\documentclass[aps,twocolumn,superscriptaddress,prb,showpacs]{revtex4-1}

\usepackage{graphicx} 
\usepackage{amsmath} 
\usepackage{epstopdf}
\usepackage[abs]{overpic} 
\usepackage{subfigure} 
\usepackage{cancel}
\usepackage{nicefrac}

\newcommand{\fcite}[1]{\footnote{Ref.~\onlinecite{#1}}}

\begin{document}

\title{\emph{Ab initio} magneto-optical spectrum of Group-IV -- Vacancy color centers in diamond}

\author{Gerg\H{o} Thiering} \affiliation{Wigner Research Centre for Physics,
Hungarian Academy of Sciences, PO Box 49, H-1525, Budapest, Hungary}
\affiliation{Department of Atomic Physics, Budapest University of Technology and
Economics, Budafoki \'ut 8., H-1111 Budapest, Hungary}

\author{Adam Gali} \email{gali.adam@wigner.mta.hu} \affiliation{Wigner Research
Centre for Physics, Hungarian Academy of Sciences, PO Box 49, H-1525, Budapest,
Hungary} \affiliation{Department of Atomic Physics, Budapest University of
Technology and Economics, Budafoki \'ut 8., H-1111 Budapest, Hungary}


\begin{abstract} 
Group-IV -- Vacancy color centers in diamond are fast emerging
qubits that can be harnessed in quantum communication and sensor applications. There is an immediate quest for understanding their magneto-optical
properties, in order to select the appropriate qubits for varying needs of
particular quantum applications. Here we present results from cutting edge \emph{ab initio} calculations about the charge state stability, zero-phonon-line energies, spin-orbit
and electron-phonon couplings for Group-IV -- Vacancy color centers. Based on the analysis of our results, we develop a novel spin Hamiltonian for these qubits which incorporates the interaction of the electron spin and orbit coupled with phonons beyond perturbation theory. Our results are
in good agreement with previous data and predict a new defect for qubit applications with thermally initialized ground state spin and long spin coherence time. 
\end{abstract} 
\maketitle

\section{Introduction}
In recent years vacancy-impurity defects in diamond have become of high interest and
are important because they show great potential in various quantum technology
applications. In particular, the spin properties of the negatively charged
silicon-vacancy [SiV($-$)] color center~\cite{elso_jel_Zaitsev, Goss1996, Clark_SiV_1.68_1995, Gali:PRB2013} with a zero-phonon line (ZPL) energy at
1.682~eV and $S=1/2$ spin has been recently studied for qubit applications~\cite{Neu2011, SiV_Hepp2014, Rogers2014PRL, Sipahigil2014, Atature2015, Pingault2014}. As this defect has inversion symmetry 
($D_{3d}$ point group) it does not directly couple to external electric field, 
and as a consequence,
SiV($-$) possesses narrow~\cite{Dietrich2014} inhomogeneous linewidth and
negligible spectral diffusion~\cite{Rogers2014PRL, Sipahigil2014}. In addition,
$\sim$70\% of the total emission occurs in ZPL emission~\cite{Neu2011}, with a
corresponding Huang-Rhys factor of 0.3. These properties are promising for
realizing solid-state sources of indistinguishable single photons for quantum communication applications~\cite{Sipahigil2014}. The fine splitting in the ground and excited levels caused by spin-orbit coupling (SOC) is harnessed to realize $\Lambda$ scheme qubit operation~\cite{Becker:NatCom2016}, however, the spin coherence time of
35~ns (see Refs.~\onlinecite{Jahnke2015, Rogers2014PRL}) is short because of the fast scattering
of the electrons between the sublevels in the ground state mediated by the
dynamic Jahn-Teller (DJT) effect even at T=4.5~K. Improvement on the spin coherence
time can be achieved by cooling down the system below T=0.5~K~\cite{Sukachev:PRL2017, Becker:PRL2018}
that suppresses the density states of the phonons that can mediate this process. 
Let us mention here that SiV($-$) has also been
proposed for optical thermometry at the nanoscale using the
temperature dependent shift of the ZPL~\cite{nguyen2017all}.
We further
note that the neutral SiV, SiV($0$), with a ZPL at 1.31~eV and $S=1$ spin~\cite{Johansson_SiV0_1.31_2011},
exhibits spin coherence time almost up to a second~\cite{rose2017observation}
and relaxation time nearly a minute~\cite{rose2017observation}. SiV($0$) is
associated with the KUL1 electron paramagnetic resonance (EPR)
center~\cite{Iakoubovskii2002PRB, SiV_Edmonds_2008, Johansson_SiV0_1.31_2011, Gali:PRB2013}. We note that SiV($0$) can be found in special diamond samples where boron and silicon doping should be simultaneously
realized during the diamond chemical vapor deposition (CVD) growth~\cite{rose2017observation}.

Inspired by the success of SiV($-$) color center, germanium, another Group-IV
element in the periodic table, was introduced into the diamond lattice, either
by CVD growth~\cite{iwasaki2015germanium, Ralchenko2015, Huler2017} or high
pressure high temperature (HPHT) synthesis~\cite{GeV_Palyanov2015,
Ekimov:2015GeV, Palyanov2017, Siyushev2017}. In all samples, a new ZPL line at
602~nm (2.06~eV) was observed in the PL spectrum, and unambiguously identified
as a Ge related center because of the isotopic shift in the ZPL
line~\cite{GeV_Palyanov2015} and its first vibronic peak~\cite{Ekimov:2015GeV},
and the anharmonicity of this peak~\cite{Ekimov2017} in the PL spectrum. Theory
predicted~\cite{Goss2005, iwasaki2015germanium} that this defect possesses the
same $D_{3d}$ symmetry like SiV does. The observed optical transitions support
this conclusion~ \cite{iwasaki2015germanium, GeV_Palyanov2015}, thus the defect
is indeed GeV($-$). The 2.06-eV line has narrow (5~nm) ZPL linewidth even at room temperature, short excited state lifetime~\cite{Bhaskar2017} (6-7~ns). 
The majority of the emission is concentrated in the ZPL 
emission~\cite{GeV_Palyanov2015},
with a Huang-Rhys factor of 0.5. Very recently, the spin relaxation and
coherence times have been observed in Ge-doped diamond samples at
$T_1\sim0.34-25$ $\mathrm{\mu s}$ and $T_2 \sim 19$
$\mathrm{ns}$~(see Ref.~\onlinecite{Siyushev2017}), respectively, at T=2~K.
We note that the signatures of GeV($0$) with $S=1$ spin have been observed in
EPR spectrum in HPHT diamond~\cite{Nadolinny2016epr, Komarovskikh2017}. This 
defect might have improved spin properties similar to those of SiV($0$). The optical signature of GeV($0$) has not yet been identified.

Simultaneously with the preparation of this paper, Sn-related PL centers have
been reported in Sn implanted diamond~\cite{tchernij2017single, Iwasaki:PRL2017},
where Sn is another Group-IV element next to germanium in the periodic table. In
particular, the 620.3-nm PL signal showed the same fine level structure in PL like
that of SiV($-$) and GeV($-$)~\cite{Iwasaki:PRL2017}. By the use of this analog,
they concluded that 620.3-nm center is associated with SnV($-$). The spin 
properties and other charge states of this defect have not been reported. Next to tin, lead is the next Group-IV element in the periodic table. To the best of our knowledge, no Pb-related color centers have been reported in diamond so far. 

Understanding the magneto-optical properties and spin coherence time of 
Group-IV -- Vacancy color centers is of immediate interest and high
importance in the fast emerging field of solid state qubits. Here we present a
systematic study on the magneto-optical properties of the Group-IV -- Vacancy
defects, including Si, Ge, Sn and Pb impurities, by means of cutting edge first
principles methods. In Sec.~\ref{sec:Methods}, we describe the first principles methodology for calculating the electronic structure, spin-orbit interaction and electron-phonon coupling of the systems. We then continue with detailed description of the results in Sec.~\ref{sec:Results} where we discuss the photostability and spin Hamiltonian of the qubits. We find that PbV color center exhibit superior spin properties over the other Group-IV -- Vacancy color centers.
Finally, we conclude the results in Sec.~\ref{sec:Conclusion}. We give additional data and derivation on the developed spin Hamiltonian in the Appendices. 

\section{Methods}
\label{sec:Methods}

We characterize point defects embedded in diamond within spin-polarized
density functional theory (DFT) as implemented in the \textsc{vasp} 5.4.1
code~\cite{Kresse:PRB1996}. Our DFT method is within Born-Oppenheimer
approximation, as ions are treated as classical particles. By varying the
positions of ions one can achieve an adiabatic potential energy surface (APES)
map of the system. The global minimum of APES defines the optimized geometry of
the system. We reach this minimum upon relaxing the atomic positions till the
force acting on every ion falls below 10$^{-2}$~eV/\AA . We embed the point
defects into a 512-atom diamond supercell, which suffices to sample the
Brillouin-zone only at the $\Gamma$-point for converged charged density. 
A relatively low energy cutoff (370~eV) for the expansion of the plane waves 
within the applied projector-augmentation-wave-method
(PAW)~\cite{Blochl:PRB1994, Blochl:PRB2000} yields converged results. We calculated the excited states
with the constrained-occupation DFT method (CDFT)~\cite{Gali:PRB2009}. We
relaxed the atomic positions by minimizing the forces acting on them in the
excited electronic state within CDFT method. The electronic structure is
calculated using HSE06 hybrid functional~\cite{Heyd03, Krukau06} within DFT.
This technique reproduces the experimental band gap and the charge transition
levels in Group-IV semiconductors within 0.1~eV accuracy \cite{Deak:PRB2010}.
This procedure also yields excellent results for the zero-phonon-line energy of
SiV($-$) center in diamond~\cite{Gali:PRB2013}. We determine the adiabatic
charge transition levels or photoionization energy thresholds as 
\begin{equation}
E(q|q+1)=E_\text{tot}^q-E_\text{tot}^{q+1}+ \Delta E^q_\text{corr}-\Delta
E^{q+1}_\text{corr} \text{,} \label{eq:transition} 
\end{equation} 
where the $\Delta E^{q}$ is the total energy correction for the point defect with 
$q$ charge by following the procedure of Lany and Zunger~\cite{Makov1995,
LanyZunger08}, while the $E^{q}_{tot}$ the total energy of the system including
the ions and electrons. We will provide these charge transition levels graphically with respect to the valence band maximum in Fig.~\ref{fig:KScharge}.

For the calculation of the phonon sideband of the PL spectrum, we calculated the vibration modes of the defects in a quasiharmonic approximation in the groundstate at high symmetry, with equal occupation of the degenerate orbitals in the band gap. We used the numerical derivatives of the forces to generate the Hessian matrix that we diagonalized to obtain the phonon frequencies and normal modes. The geometry is preoptimized with the very strict force criterion of $10^{-4}$~eV/\AA for the vibration calculations. We consistently applied the computationally powerful Perdew-Burke-Ernzerhof (PBE) functional~\cite{PBE} in this procedure that reproduces the calculated HSE06 phonon spectrum of the perfect diamond within 5~cm$^{-1}$ and the quasilocal modes of SiV($-$) within 2~cm$^{-1}$.

We determined the spin-orbit coupling (SOC) in non-collinear approach as
implemented in \textsc{VASP}~5.4.1 for the negatively charged 
Group-IV -- Vacancy centers, or briefly, XV centers. We set the quantization axis 
of the spin along $\langle111\rangle$ direction, the $C_3$ rotation axis of the 
XV point defects. We determined the SOC parameters by HSE06 DFT functional, which has generally provided accurate results for the spin-orbit splitting of nitrogen-vacancy center in diamond~\cite{Thiering2017SOC}. SOC is a small
perturbation to the electronic structure of the system, thus we fixed the atomic positions in the high symmetry $D_{3d}$
configuration as obtained from the previous spin-polarized DFT geometry
relaxations. As was reported in our previous study~\cite{Gali:PRB2013} for SiV center and is also shown in Sec.~\ref{sec:Methods} for other XV centers, a double degenerate $e_{g\{x,y\}}$ level appears in the gap, which will be occupied by three electrons in the negatively charged state, which can be treated as a single hole on this state. After applying the SOC on the system, 
$e_{g\pm}=\frac{1}{\sqrt{2}}
\left(e_{gx} \pm i e_{gy}\right)$ states split by $\lambda_0$ (see Fig.~\ref{fig:KScharge} for the electronic structure) which comes from the $z$ component of SOC. By using CDFT
procedure it is feasible to introduce the hole either on the $e_{g+}$ or 
$e_{g-}$ state, and
the calculated total energy difference is the strength of spin-orbit coupling.
The Hamiltonian of SOC coupling is the following: 
\begin{widetext}
\begin{equation}
\hat{H}_{SOC}= -\lambda_{0}\hat{L}_{z}\hat{S}_{z}= -\frac{\lambda_{0}}{2}\left[\left|e_{g+}^{\uparrow}\right\rangle
\left\langle e_{g+}^{\uparrow}\right|+\left|e_{g-}^{\downarrow}\right\rangle
\left\langle e_{g-}^{\downarrow}\right|-\left|e_{g-}^{\uparrow}\right\rangle
\left\langle e_{g-}^{\uparrow}\right|-\left|e_{g+}^{\downarrow}\right\rangle
\left\langle e_{g+}^{\downarrow}\right|\right] \label{eq:SOC} 
\end{equation}
\end{widetext} 
where $\hat{L}_z$ is the effective orbital moment operator
($L=1$) of the electron, while $\hat{S}_z$ is the electronic spin. We find
that half of this total energy difference is equal within $10^{-7}$~eV to
the split of $e_{g+}$ and $e_{g-}$ Kohn-Sham levels when these two states are
occupied by half-half electrons. Thus, the strength of SOC can be calculated by
the half-half occupation of the $e$ states with following the SOC splitting of
these $e_{g}$ states. The negative sign of $\lambda_0$ accounts for the fact that $e_{g}$ particle is a hole and not an electron. The SOC Hamiltonian \eqref{eq:SOC} can be expressed with the Pauli matrices
$\hat{H}_{SOC}=\frac{\lambda_{0}}{2}\sigma_{y}$ where 
\begin{equation}
\sigma_{y}=\begin{pmatrix} & -i\\
	i & 
\end{pmatrix}\text{,}
\label{eq:Pauli-y} 
\end{equation}
that represent the $e_{g\pm}$ electron states.

We note that the dynamic Jahn-Teller (DJT) effect quenches the orbital moment, at
least partially, known as the Ham effect~\cite{Ham_1965, bersuker2006, bersuker2012vibronic}. The intrinsic
$\lambda_{0}$ that we calculate directly from Kohn-Sham orbitals, is
severely reduced by the Ham reduction factor $p$, thus $\lambda_\text{Ham}=p\lambda_{0}$ reduced value is observed in the experiments. 

We now discuss two cases: (i) the electron-phonon coupling manifested as DJT effect is significantly larger than spin-orbit coupling, so one can solve first the electron-phonon system and then calculate the spin-orbit energies as a first perturbation acting on the resultant vibronic wavefunctions; (ii) the electron-phonon coupling and spin-orbit coupling are in the same order of magnitude, so the orbital, spin, and phonon degrees of freedom of the wavefunction are strongly coupled which requires the exact diagonalization of the sum of spin-orbit and electron-phonon Hamiltonians, respectively. We call the first case as Ham reduction factor solution whereas the second case the exact diagonalization procedure.

We first describe the case (i) where we define the electron-phonon Hamiltonian caused by DJT effect. The DJT effect is 
the interaction of an $E_g$ type quasilocalized vibration mode with the $e_g$
electron orbital, known as $E\times e$ DJT system. The Hamiltonian of such system 
is the following: 
\begin{widetext}
\begin{equation}
\hat{H}_{DJT}=\hbar\omega_{e}\left(a_{x}^{\dagger}a_{x}+a_{y}^{\dagger}a_{y}+1\right)+F\left(\hat{x}\sigma_{z}-\hat{y}\sigma_{x}\right)+G\left[\left(\hat{x}^{2}-\hat{y}^{2}\right)\sigma_{z}+2\hat{x}\hat{y}\sigma_{x}\right]	 
\text{,} \label{eq:DJT} 
\end{equation} 
\end{widetext} 
where the $a_{x,y}^{\dagger},
a_{x,y}$ are the creation and annihilation operators of the $E_g$ vibration
mode, respectively. The system is a twodimensional harmonic oscillator with
frequency $\omega_e$, where the terms labeled by $F$ and $G$ parameters are the corresponding linear and quadratic part of DJT. The $\sigma_i$ operators are 
twodimensional Pauli matrices 
\begin{equation}
\sigma_{z}=\begin{pmatrix}1\\
	& -1
\end{pmatrix}\qquad\sigma_{x}=\begin{pmatrix} & 1\\
1
\end{pmatrix}
\label{eq:Pauli} 
\end{equation}
representing the $e_{gx}$ and $e_{gy}$ electrons,
respectively. $F$ and $G$ parameters are the values describing the Jahn-Teller
distortion through $(\hat{x},\hat{y})=\frac{1}{\sqrt{2}}\left(a_{(x,y)}^{\dagger}+a_{(x,y)}\right)$ operators. $F$ and $G$ parameters can be easily derived after the 
APES of the DJT system is determined. The energy gain from the symmetry 
distortion (see Chapter 3.2.\ in Ref.~\onlinecite{bersuker2006}) to one of the three global minima is 
$E_\text{JT}=\frac{F^2}{2\left(\hbar \omega_e - 2G \right)}$, and the barrier
energy separating the these three minima is $\delta_\text{JT} = 
\frac{4 E_\text{JT} G
}{\hbar \omega_e+2G}$. For SiV($-$), GeV($-$), and with less extent, for
SnV($-$) in the ground state, the energy gain from DJT, i.e., $E_\text{JT}$ is orders of magnitude larger than the energy of SOC coupling. Thus, SOC can be evaluated as a perturbation on the DJT groundstate wavefunction, where the electrons and vibrations are entangled. We determine the eigenvalues of the Hamiltonian in Eq.~\eqref{eq:DJT}
numerically with the following series of expansion, 
\begin{equation}
\left|\tilde{\Psi}_{\pm}\right\rangle =\sum_{n,m}\left[c_{nm}\left|e_{g\pm}\right\rangle
\left|n,m\right\rangle +d_{nm}\left|e_{g\mp}\right\rangle \left|n,m\right\rangle
\right] \text{.} \label{eq:DJTseries} 
\end{equation} 
We then express the SOC splitting within perturbation theory as 
$\ensuremath{\lambda_{\mathrm{Ham}}=2\left|\left\langle \tilde{\Psi}_{\pm}\right|\hat{H}_{SOC}\left|\tilde{\Psi}_{\pm}\right\rangle \right|=p \lambda_{0}}$.
We limit the series expansion of the twodimensional harmonic oscillator 
$\left|n,m\right\rangle$ up to ten quanta, thus $n+m \leq 10$. From this expansion, the $p$ reduction factor can be expressed as $p=\sum_{n,m} \left( c_{nm}^2 - d_{nm}^2 \right)$ with $c_{nm}$ and $d_{nm}$ expansion coefficients (see Appendix \ref{sec:derHamfactor} for derivation) .

We next describe case (ii) where DJT and SOC energies are comparable, so the corresponding Hamiltonians in Eqs.~\eqref{eq:DJT} and \eqref{eq:SOC} should be added and solved simultaneously with coupled orbital, spin and phonon wavefunctions. Here we expanded the polaronic wavefunction $\tilde{\Psi}_{\Gamma}$ with the spin degrees of freedom as
\begin{equation}
\left|\tilde{\Psi}_{\Gamma}\right\rangle =\sum_{n,m}\left[c_{nm}^{\chi}\left|e_{g\pm}\right\rangle \left|n,m\right\rangle \left|\chi\right\rangle +d_{nm}^{\chi}\left|e_{g\mp}\right\rangle \left|n,m\right\rangle \left|\chi\right\rangle \right]
\label{eq:DJTSOCseries}
\end{equation} 
and then directly diagonalize the Hamiltonian including the SOC and DJT effects simultaneously, where $\chi$ can be either $\uparrow$ or $\downarrow$ spin state. 
This solution represents a coupling between spins and phonons and goes \emph{beyond} the perturbation theory of SOC acting on the polaronic wavefunctions. Here, the subindex $\Gamma$ refers to the total angular momentum of the wavefunction which is either $\nicefrac{3}{2}$ or $\nicefrac{1}{2}$ (see also Appendix~\ref{sec:dereffHam}). We note that the simultaneous treatment of SOC and DJT in $E\times e$ JT systems has been only considered for very small molecules at \emph{ab initio} level in the literature so far~\cite{Aubock2008, mondal2012ab}. However, extensive \emph{ab initio} study on point defects in solids, with a model consisting of hundreds of atoms, has not yet been performed, to the best of our knowledge. As we will show below our methodology is able to reproduce previous experimental data.

\section{Results}
\label{sec:Results} 

\begin{figure*}[ht] 
\includegraphics[width=\textwidth]{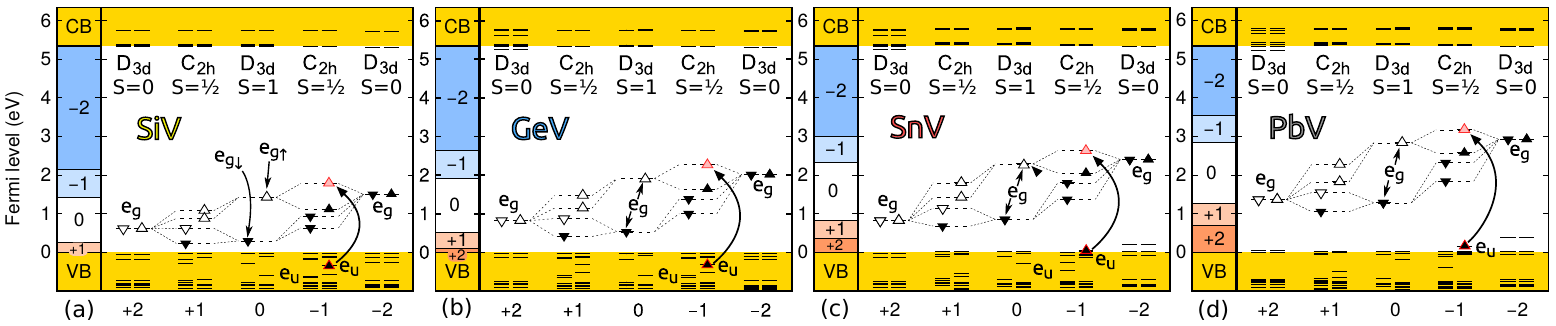} 
\caption{\label{fig:KScharge}Kohn-Sham single particle levels and charge transition 
levels of XV defects for X=Si, Ge, Sn, and Pb. We note that the effect of spin-orbit interaction is not included here. We represent the $\uparrow$ ($\downarrow$) spin channel with triangles pointing upwards (downwards). The filled (empty) triangles depicts occupied one electron (hole) orbitals. In the 
($+$) and ($-$) charge states, the symmetry is lowered from $D_{3d}$ to $C_{2h}$ within adiabatic potential energy surface in the DFT calculations. We show the charge stability window for the given charge state referenced to the valence band maximum of diamond at the left panel in each figure. The excitation processes are also shown for XV($-$) color centers. The $e_u$ orbital is resonant with the valence band that pops up in the excited electron configuration. }
\end{figure*}

All the Group-IV impurities reside in the symmetric
split-vacancy configuration in diamond according to our calculations 
that results in a $D_{3d}$ symmetry. The split-vacancy configuration may be
labeled as V-X-V which implies that the 'X' impurity atom lies at an interstitial
position, more precisely, at the inversion point of the diamond lattice halfway
between
two adjacent vacancies, or briefly, in divacancy. Nevertheless, the quantum optics
groups labeled these defects by XV in the literature, thus we consequently used this
notation in the Introduction and the rest of the paper. According to this description,
this type of defects
exhibits an inversion symmetry which adds a parity to the wavefunctions, odd
(ungerade) and even (gerade), labeled by $u$ and $g$, respectively. One can construct the possible single electron orbitals from the defect-molecule model. There are 6 carbon dangling bonds pointing towards the impurity atom, from which 6 orbitals emerge in $D_{3d}$ symmetry: $a_{1g}+a_{2u}+e_{u}+e_{g}$. The corresponding orbitals are filled with 10 valence electrons where 6 electrons comes from the dangling bonds and 4 electrons from the impurity atom.
As discussed previously for SiV defect~\cite{Gali:PRB2013}, only the $e_g$ orbital appears in
the gap which is filled by two electrons with parallel spins in the neutral
charge state (see also Fig.~\ref{fig:KScharge}) whereas the $a_{1g}$ and $a_{2u}$ levels fall inside the valence band and $e_u$ level is resonant with it. The same $e_g$ orbital occurs in the gap
for the other XV defects. The in-gap optical excitation involves an $e_u$
orbital resonant with the valence band (VB) that pops up in the gap in the
excited state electronic configuration. Special bound exciton states might occur
between in-gap $\leftrightarrow$ band edge optical excitation that might also
lead to photoionization. The electronic structure of the neutral XV
complexes implies that charge states from $(2+)$ to $(2-)$ can exist depending
on the position of the (quasi) Fermi-level in diamond. Here, we focus our study
on the $(-)$ charge state (coherent dark states in the prototype SiV($-$) qubit~\cite{Pingault2014}) and the neutral charge state (optical spin polarization with long spin coherence 
time of prototype SiV($0$)~\cite{rose2017observation}), i.e., their charge state stability. We provide trends for 
the magneto-optical properties of the negatively charged XV complexes.

\subsection{Charge state stability of the XV defects}
\begin{figure}[h]
\includegraphics[width=225pt]{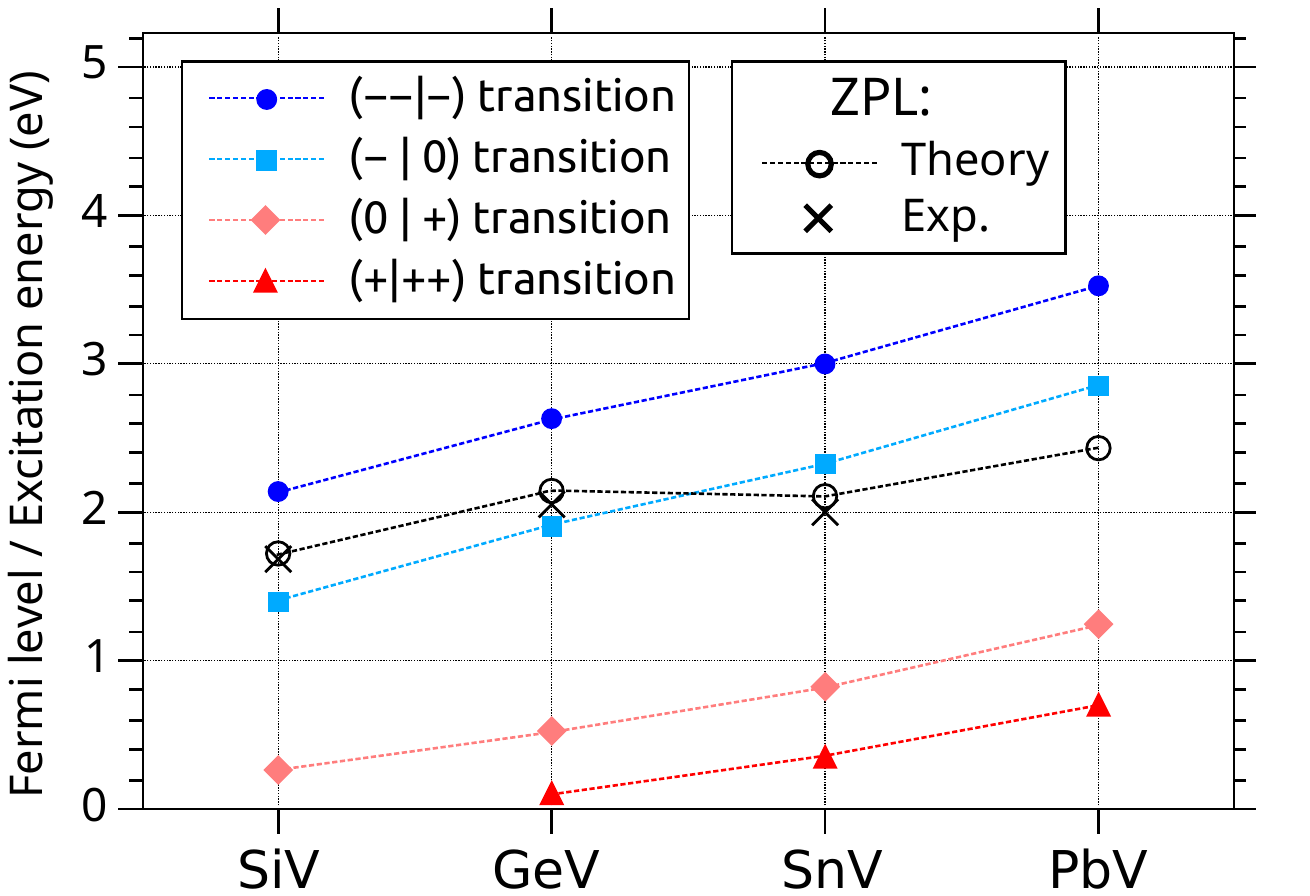}
\caption{\label{fig:trends}Adiabatic photoionization of XV defects (see Eq.~\ref{eq:transition}) and excitation energies for XV($-$) centers (see Sec.~\ref{ssec:PL}). All the charge transition levels are referenced to the valence band maximum aligned to zero. The conduction band minimum is at 5.4~eV. We note that these energies do not contain spin-orbit energy corrections. See text for explanation.}	
\end{figure}

A general trend in the electronic structure of XV defects is that the
$e_g$ level shifts up with increasing atomic number of Group-IV impurity atom (see Fig.~\ref{fig:KScharge}) which has a consequence on the formation energies and the corresponding adiabatic charge transition levels too. In order to readily see the trends, the calculated $(-)$$\rightarrow$$(0)$ and $(0)$$\rightarrow$$(-)$ transition 
energies are also plotted in Fig.~\ref{fig:trends}, where the first and second 
transition is associated with promoting an electron from the defect level to the 
conduction band and from the valence band to defect level, respectively, but the 
plot depicts the charge transition levels with respect to the valence band maximum. 
The stability window of $(-)$ state shifts up in the gap with increasing atomic number of Group-IV impurity atom. SnV($-$) and PbV($-$) can only be stable by providing substitutional nitrogen donors in the diamond sample. The photostability of 
PbV($-$) requires special attention as the $(-)$ state can be converted to $(0)$ by illuminating the sample in the visible region, by $\sim$2.6~eV, that it is close to its ZPL energy of 
about 2.4~eV (see Table~\ref{tab:ZPLs}). On the other hand, photoexcitation in the visible region may convert $(-)$ state to $(2-)$ charge state for SiV, GeV and 
SnV defects. For isolated SiV($-$), ultraviolet (UV) light would be needed to reionize $(2-)$ to $(-)$ by single photon absorption that is difficult to realize in the experiments. On the other hand, violet and blue illumination can reionize GeV($2-$) and SnV($2-$) to GeV($-$) and SnV($-$), respectively. The ($2-$) charge state is a spin singlet, nevertheless, a shelving triplet bound exciton state might exist that can be accessed by optical pumping at $\sim$2.4~eV and $\sim$1.8~eV for SnV($2-$) and PbV($2-$), respectively. The stability window of $S=1$ $(0)$ state of GeV, SnV and PbV also shifts up in the gap with increasing the atomic number of the Group-IV impurity atom.  

One can conclude that SnV($-$) defect should be very photostable whereas isolated SiV is trapped in the $(2-)$ charge state once SiV($-$) has been photoionized into
that state. On the other hand, there is a small energy margin between the
calculated neutral excitation energy ($\sim$2.43~eV) and $(-)$$\rightarrow$$(0)$ transition energy ($\sim$2.6~eV) for PbV($-$), thus this defect can be photoionized into the $(0)$ charge state by blue illumination into the phonon sideband.

We note that the neutral XV($0$) with $S=1$ ground state may also act like a qubit, with presumably long electron spin coherence time in good quality of diamond. Our calculations show that SnV($0$) and PbV($0$) defects can be engineered into typical diamond samples where nitrogen contamination occurs, in contrast to the case of SiV($0$) defect, which requires boron doping of diamond. Formation of GeV($0$) requires very low nitrogen concentration or compensation of nitrogen donors by acceptors.

\subsection{Photoluminescence of XV($-$) defects}
\label{ssec:PL}
\begin{figure*}[ht] 
\includegraphics[width=480pt]{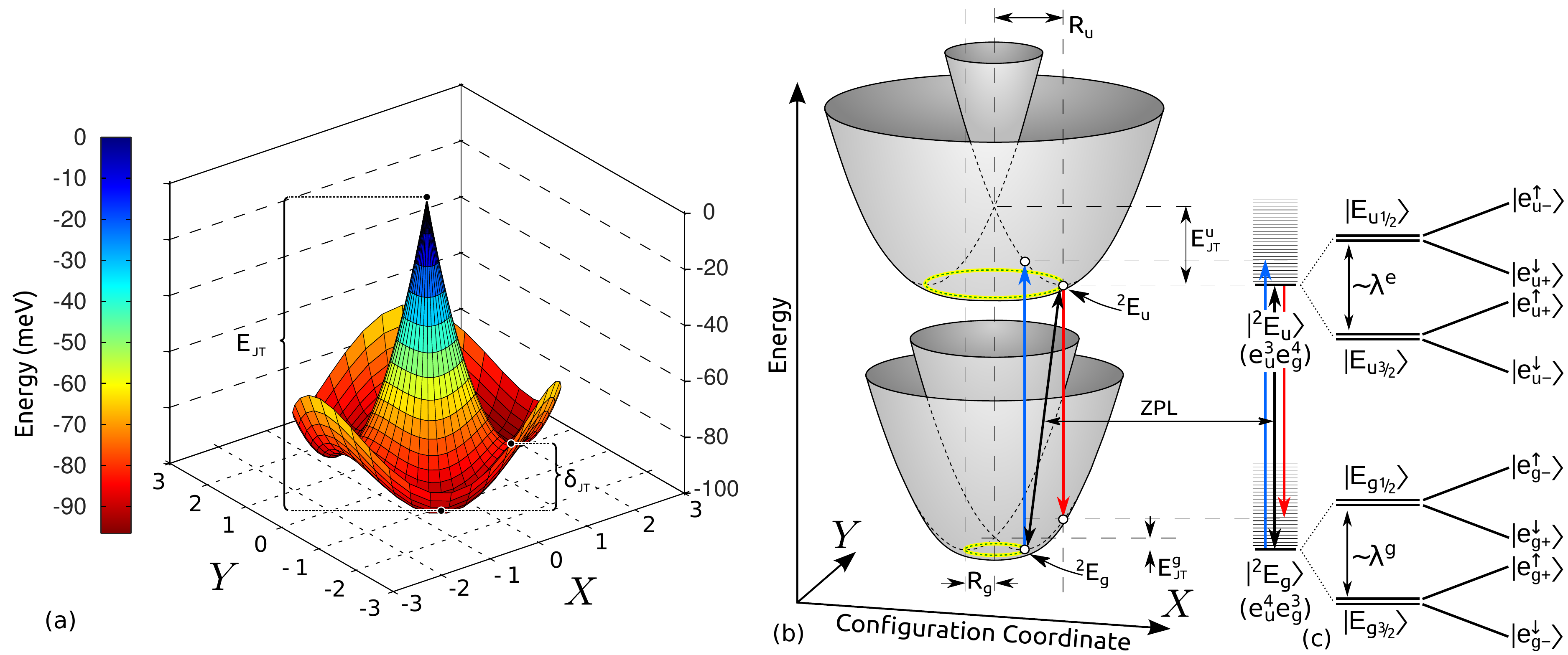} 
\caption{\label{fig:KSTeller}(a) Jahn-Teller adiabatic potential energy surface (APES) of the optically excited $^2E_u$ state in PbV($-$). (b) Jahn-Teller APES of 
XV($-$) defects as a function of $e_x$ and $e_y$ phonon distortions, with quadratic JT effects neglected, which results in an APES with axial symmetry. In the ground state ($g$) the global APES minimum can be found at $R_g$ distance ($C_{2h}$ distortion) from the high symmetry $D_{3d}$ geometry while this is $R_u$ for the $u$ excited state. $E_\text{JT}$ is the Jahn-Teller energy is the energy difference between the $D_{3d}$ and $C_{2h}$ configurations. $\delta_\text{JT}$ energy barrier between the three local minima of C$_{2h}$ configurations (few meVs) is not shown here for the sake of clarity. On the right panel (c), we schematically depict the phonon sideband of the ground and excited state where blue (red) lines represent the excitation (luminescence) of the system into the phonon sideband, whereas the black line defines the zero-phonon line transition (ZPL).
The spin-orbit coupling ($\lambda$) split $^2E_g$ and $^2E_u$ ZPL states apart into $E_{g\nicefrac{1}{2}}$,$E_{g\nicefrac{3}{2}}$,$E_{u\nicefrac{1}{2}}$,$E_{u\nicefrac{3}{2}}$ Kramers doublets where $\nicefrac{3}{2}$ and $\nicefrac{1}{2}$ refers to an effective total angular momentum of the electronic states. The doublets may further split into individual $e_{g,u\pm}^{\uparrow/\downarrow}$ states under external static magnetic field.} 
\end{figure*}

The photoluminescence can be described by spontaneous emission from the optically lowest energy excited state to the groundstate.
The lowest energy excited state can be understood as promoting an electron from the $e_u$ orbital to the $e_g$ orbital in XV($-$) defects. As a consequence, the excited state is $^2E_u$ and the ground state is $^2E_g$ in the negatively charged state~(see Ref.~\onlinecite{Gali:PRB2013} and Fig.~\ref{fig:KScharge}). Both $^2E_u$ and $^2E_g$ states are dynamic Jahn-Teller systems~\cite{Gali:PRB2013, SiV_Hepp2014, Jelezko_2014}. We find a general trend in the calculated Jahn-Teller energy, $E_\text{JT}$, as a function of the atomic number of the Group-IV impurity atom, where $E_\text{JT}$ is defined as the total energy difference in the high symmetry $D_{3d}$ geometry and the lowest energy $C_{2h}$ geometry in the adiabatic potential energy surface (APES) (see Fig.~\ref{fig:KSTeller}). The ZPL energies are calculated by taking the lowest APES energy in $C_{2h}$ symmetry both in the ground and excited state (see Table~\ref{tab:ZPLs}) that we call here an "average" method. Interestingly, the calculations do not predict linearly increasing trend in the ZPL energies by increasing the atomic number of the Group-IV impurity but the ZPL of SnV($-$) should be smaller than that of GeV($-$). This agrees well with the very recent experimental data~\cite{Iwasaki:PRL2017}.
\begin{table}[h] 
\caption{\label{tab:ZPLs}The calculated zero-phonon-line (ZPL) energies and Huang-Rhys factors ($S$) for XV($-$) defects are given and compared to the experimental data (ZPL$_\text{exp}$ and $S_\text{exp}$). Here we provide the average ZPL values in C$_{2h}$ symmetry [ZPL($C_{2h}$)] and within exact calculation of dynamic Jahn-Teller (DJT) effect together with spin-orbit coupling [ZPL(SOC)]. The latter more accurate method brings our results closer to the experimental values. See text for explanation.} 
\begin{ruledtabular}
\begin{tabular}{lcccc}
                                    & SiV   & GeV  & SnV  & PbV \\ \hline 
 ZPL($C_{2h}$) (eV)      & 1.72 & 2.15  & 2.11 & 2.45 \\
 ZPL(SOC) (eV)             & 1.70 & 2.12  & 2.09 & 2.40 \\
 $S$                             & 0.27 & 0.50  & 0.89 & 1.60 \\
 ZPL$_\text{exp}$ (eV)& 1.68\fcite{SiV_Zaitsev_1981} & 2.06\fcite{Ekimov:2015GeV, GeV_Palyanov2015, GeV_Palyanov2016}  & 2.00\footnote{Ref.~\onlinecite{Iwasaki:PRL2017} \label{foot:SnZPL}} & n.a. \\
 $S_\text{exp}$           & 0.24\fcite{SiV_Collins1994, SiV_Neu2011} & 0.50\fcite{GeV_Palyanov2015}  & 0.89\textsuperscript{\ref{foot:Si}} & n.a. \\
\end{tabular} 
\end{ruledtabular} 
\end{table}

Regarding XV($-$) qubits or qubit candidates, we show the calculated average ZPL energies compared to the experimental ones when available (see Table~\ref{tab:ZPLs}). The calculated average ZPL energies, that do not contain the zero-point energies and spin-orbit couplings in the ground and excited state, somewhat overestimate the experimental ones. By an accurate calculation of the zero-point energies within our DJT treatment together with the spin-orbit coupling, brings the computed ZPL energies closer to the experimental ones, with respect to the average ZPL energies. Nevertheless, both methods are accurate within 0.1~eV (see 
Fig.~\ref{fig:trends}). This gives us confidence that the calculated ZPL energy of PbV($-$) is well predicted. As can be read out from Table~\ref{tab:ZPLs} the nature of ZPL transition involves polaronic states together with spin-orbit effects (see Appendix~\ref{sec:dereffHam}). Thus, perturbation effects, e.g.\ strain or temperature, on ZPL energies should involve the complex analysis of this coupled spin-orbit-phonon system. The calculation of spin-orbit coupling and electron-phonon coupling will be described in the next sections.

Because the excited states can be well described by promoting an electron from the $e_u$ orbital to the $e_g$ orbital according to our CDFT calculations, therefore, the optical transition dipole moment for XV($-$) defects can be well approximated by calculating the optical dipole moment $\mu$ between these Kohn-Sham states. The radiative lifetime of these color centers then can be 
calculated~\cite{Weisskopf:1930} as 
\begin{equation}
\label{eq:lifetime}
\tau = \frac{n \omega^3 |\mu|^2}{3\pi \epsilon_0}\text{,}
\end{equation}
where $n$ is the refractive index and $\hbar\omega$ is the excitation energy. The calculated radiative lifetimes are listed and compared to the observed PL lifetime of the XV($-$) defects in Table~\ref{tab:lifetimes}. A general trend is that the computed radiative lifetime somewhat decreases with increasing atomic number of the impurity atom but basically they are all in the same order of magnitude. The predicted short radiative lifetime ($\approx$3~ns) of PbV($-$) center is favorable for quantum emitter applications. We note that the observed PL lifetime of SiV($-$) is significantly shorter than its computed radiative lifetime. We attribute this effect to the strong non-radiative processes which we tentatively assign to the ionization process that competes with the neutral excitation process (see also a recent photoluminescence excitation observation in Ref.~\onlinecite{Hassler:2018NJP}). The calculated $(2-|-)$ charge transition level of SiV($-$) at $\approx$2.05~eV is very close to the energy of the phonon sidebands of neutral excitation whereas the energy between these two increases with increasing atomic number of the impurity atom that should suppress this type of non-radiative processes. In addition, the optical gap also significantly larger for GeV($-$), SnV($-$), and PbV($-$) than that of SiV($-$), which also suppresses the direct non-radiative decay induced by phonons in the former color centers.    
\begin{table}
\caption{\label{tab:lifetimes}The calculated radiative lifetimes ($\tau_\text{rad}$) versus the observed photoluminescence lifetimes ($\tau_\text{PL}$) for XV($-$) color centers at cryogenic temperatures. We used the experimental ZPL energy where available in the calculation of $\tau_\text{rad}$. We note that $\tau_\text{PL}$ involves both radiative and non-radiative processes.}
\begin{ruledtabular}
\begin{tabular}{lcccc}
                                    & SiV($-$) & GeV($-$) & SnV($-$) & PbV($-$) \\
$\tau_\text{rad}$ (ns)  &  12.13    & 6.62        &  5.49       & 2.88   \\ 
$\tau_\text{PL}$ (ns)   & 1.72\footnote{Ref.~\onlinecite{Rogers:2014NatCom}}      
                                    &  $\sim$6\footnote{Ref.~\onlinecite{Bhaskar2017}} 
                                   &  $\sim$5\footnote{Ref.~\onlinecite{Bhaskar2017}} 
                                   &  n.a. \\
\end{tabular}
\end{ruledtabular}
\end{table}

\begin{figure*}[ht] \includegraphics[width=\textwidth]{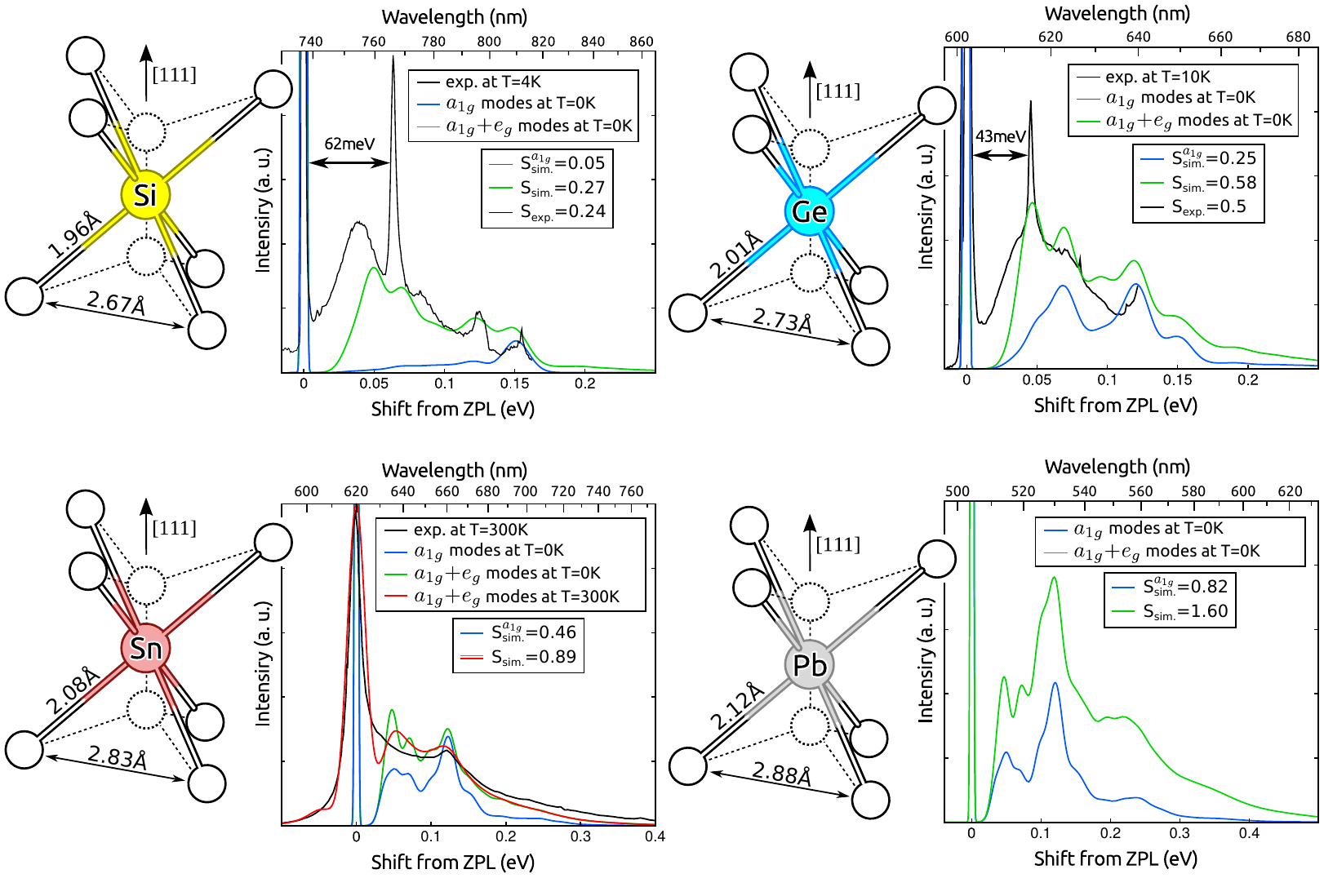} 
\caption{\label{fig:XVm1_HR} Experimental and theoretically predicted luminescence lineshapes of XV($-$) complexes. The red-green-blue curves are theoretical predictions with HR theory, while the black lines are the actual experimental data for SiV($-$) and GeV($-$). The blue curves contain only the effect 
of totally symmetric $a_{1g}$ phonons whereas the green curves add the contribution of $e_g$ phonons that appear due to the Jahn-Teller nature of the electronic states. The experimental curve for SiV($-$) and GeV($-$) is taken from Refs.~\onlinecite{Dietrich2014} and \onlinecite{Ekimov:2015GeV}, respectively. For SnV($-$), we use the 620.3-nm PL spectrum that has been recently recorded in Sn-implanted diamond sample at room temperature (Ref.~\onlinecite{Iwasaki:PRL2017}). We show the room temperature theoretical spectrum for SnV in the red plot. We set the linewidth of ZPL to match with the experimental
data as our present method cannot determine the ZPL broadening. $S$ values are the calculated Huang-Rhys factors taken from our simulations (sim.) or experiments (exp).}  
\end{figure*}

The phonon sideband in the PL spectrum is determined within the Huang-Rhys (HR) theory~(see the original theory in Ref.~\onlinecite{Huang1950} and our implementation in Ref.~\onlinecite{Gali:2016Ncomm} that is based on Ref.~\onlinecite{Alkauskas2014}). We
calculate the HR spectra between two statically distorted Jahn-Teller structures
as depicted in Fig.~\ref{fig:KSTeller}. By this way, we can take into account the contribution of the $e_g$ phonons in the PL spectrum that are responsible for the
most intense phonon sidebands for SiV($-$) and GeV($-$) defects. 
These $e_g$ phonons are bulk-like in nature and do not localize on the defect. The
participation of these $e_g$ phonons in the phonon sideband is a consequence of
the dynamic Jahn-Teller nature of the groundstate and excited state. We note that
quasilocal vibration modes are also visible as relatively sharp features 
in the PL spectrum of SiV($-$) and GeV($-$)
at about 62~meV and 43~meV, respectively, 
that are not reproduced by our method.
Based on the calculated vibration spectrum of SiV($-$) defect in our previous
study~\cite{Londero:arxiv2016},
we associate this feature with the $e_u$ quasilocal vibration modes of the defects
that involve the ($x$, $y$) motions of the impurity atom. Principally, the usual
Franck-Condon approximation on the luminescence of polyatomic systems does
not allow the participation of \emph{ungerade} modes in the PL process, thus
the observations might be explained by invoking the Herzberg-Teller effect that
goes beyond the Franck-Condon approximation
(see Ref.~\onlinecite{Londero:arxiv2016} for details). This issue is beyond the
scope of the present study, and we rather focus on the general trends in the PL
spectra. As can be seen in Fig.~\ref{fig:XVm1_HR} the contribution of phonons to the PL spectrum increases with increasing atomic number of Group-IV impurity. In particular, the contribution of $a_{1g}$ phonons significantly increases because the geometry change between the ground and excited state's geometries increases, as measured by the Huang-Rhys factor $S$. This can be understood by the size of the heavy atoms that cannot readily be accommodated by the divacancy of diamond and they start to substantially distort the diamond lattice. We note that this trend is disadvantageous for creating very efficient single photon sources emitting light dominantly in the zero-phonon emission, nevertheless, the calculated $S=1.60$ factor for PbV($-$) is still much smaller than that $S \approx 3.5$ for NV($-$) center in diamond (see Ref.~\onlinecite{Thiering2017SOC} for detailed analysis). 
 
\subsection{Effective spin-orbit coupling in XV($-$) defects} 
A splitting occurs in both the $^2E_u$ and $^2E_g$ levels due to spin-orbit coupling (SOC) between the $S=1/2$ electron spin and the double degenerate orbital forming two Kramers doublets that we call zero-field-splitting (ZFS) where zero-field refers to zero external magnetic field. In DJT systems, the spin-orbit coupling can reduce the effective SOC by the $p$ Ham reduction factor~\cite{Ham_1965, Ham_1968}. We use here the same \emph{ab initio} theory to calculate the $p$ factor as we demonstrated for the $^3E$ excited state of the negatively charged nitrogen-vacancy center in diamond~\cite{Thiering2017SOC}. This requires to calculate the full APES in the corresponding electronic state as depicted in Fig.~\ref{fig:KSTeller}. Here we note that $E_\text{JT}$ decreases with increasing atomic number of Group-IV impurity (see Table~\ref{tab:SOC}). This results in smaller damping of SOC with increasing atomic number of Group-IV impurity. The intrinsic spin-orbit coupling is determined as described in Sec.~\ref{sec:Methods}. Accurate calculation of spin-orbit coupling requires scaling method in giant supercells (see Ref.~\onlinecite{Thiering2017SOC} and Appendix~\ref{sec:SOCscale}). The results are summarized in Table~\ref{tab:SOC}.
\begin{table*}[htbp] 
\caption{\label{tab:SOC}The calculated basic parameters of the APES such as $E_\text{JT}$ Jahn-Teller energy, $\delta_\text{JT}$ barrier energy and the $\hbar \omega_e$ energy of the effective $e_g$ phonon driving the Jahn-Teller effect are shown as well as the calculated $\lambda_0$ intrinsic spin-orbit coupling and $p$ Ham reduction factor, and the deduced $\lambda_\text{Ham}=p\lambda_0$ effective spin-orbit coupling for the $^2E_g$ optically ground state and $^2E_u$ excited state of XV($-$) defects. The calculated zero-field splitting ($\lambda$) is for comparison to the experimental value ($\lambda_\text{exp}$). We determine the calculated $\lambda$ values beyond the simple Ham reduction theory where we treat the SOC and DJT Hamiltonians simultaneously (see Fig.~\ref{fig:SOC} for graphical interpretation). } 
\begin{ruledtabular}
\begin{tabular}{lcccccccc}
 system & $\lambda_0$ (meV, GHz) & $E_\text{JT}$ (meV) & $\delta_\text{JT}$ (meV) & 
 $\hbar\omega$ (meV) & p & $\lambda_\text{Ham}$ (GHz) & $\lambda$ (GHz) & $\lambda_\text{exp}$ (GHz)\\ \hline 
SiV($^2E_g$) & 0.82, 198  & 42.3 & 3.0 & 85.2 & 0.308 & 61.0 & 61.0 & 50\footnote{Ref.~\onlinecite{SiV_Hepp2014}   \label{foot:Si}}  \\ 
GeV($^2E_g$) & 2.20, 532  & 30.1 & 2.0 & 82.2 & 0.390 & 207  & 207  & 181\footnote{Ref.~\onlinecite{Ekimov:2015GeV} \label{foot:Ge}}\\ 
SnV($^2E_g$) & 8.28, 2001 & 21.6 & 1.6 & 79.4 & 0.472 & 946  & 945  & 850\footnote{Ref.~\onlinecite{Iwasaki:PRL2017} \label{foot:Sn}}\\ 
PbV($^2E_g$) & 34.6, 8360 & 15.6 & 0.6 & 74.9 & 0.540 & 4514 & 4385 & n.a. \\ \hline 
SiV($^2E_u$)& 6.96, 1680 & 78.5 & 2.7  & 73.5 & 0.128 & 215  & 215  & 260\textsuperscript{\ref{foot:Si}}\\ 
GeV($^2E_u$)& 36.1, 8720 & 85.7 & 5.4  & 73.0 & 0.113 & 987  & 989  & 1120\textsuperscript{\ref{foot:Ge}} \\ 
SnV($^2E_u$)& 96.8, 23200 & 83.1 & 6.8  & 75.6 & 0.125 & 2897 & 2925 & 3000\textsuperscript{\ref{foot:Sn}} \\ 
PbV($^2E_u$)& 245,  59300 & 91.6 & 12.3 & 78.6 & 0.119 & 7051 & 6920 & n.a.\\ 
\end{tabular} 
\end{ruledtabular} 
\end{table*}

The general trend is that the $\lambda_0$ rapidly grows with increasing atomic number of 
Group-IV impurity. In the $^2E_g$ ground state, there is a turning point for 
PbV($-$) defect where $\lambda_0$ is greater than $E_\text{JT}$, thus SOC is not 
a small perturbation w.r.t.\ electron-phonon coupling but the electron-phonon 
Hamiltonian has to be parallel diagonalized together with the spin-orbit 
Hamiltonian (see Sec.~\ref{sec:Methods} and Appendix~\ref{sec:dereffHam}). 
As a consequence, the estimated ZFS between the 
sublevels of $^2E_g$ ground state is around 18.7~meV with the Ham reduction scheme ($\lambda_\text{Ham}$), and 18.1~meV ($\lambda$) with the exact diagonalization (see Tab.~\ref{tab:SOC} for details). The $p$ Ham reduction parameter also increases with increasing atomic number of Group-IV impurity because $E_\text{JT}$ decreases whereas the vacancy related $\omega_e$ vibration modes are relatively insensitive to the type of Group-IV impurity atom.

\begin{figure*}[ht] \includegraphics[width=\textwidth]{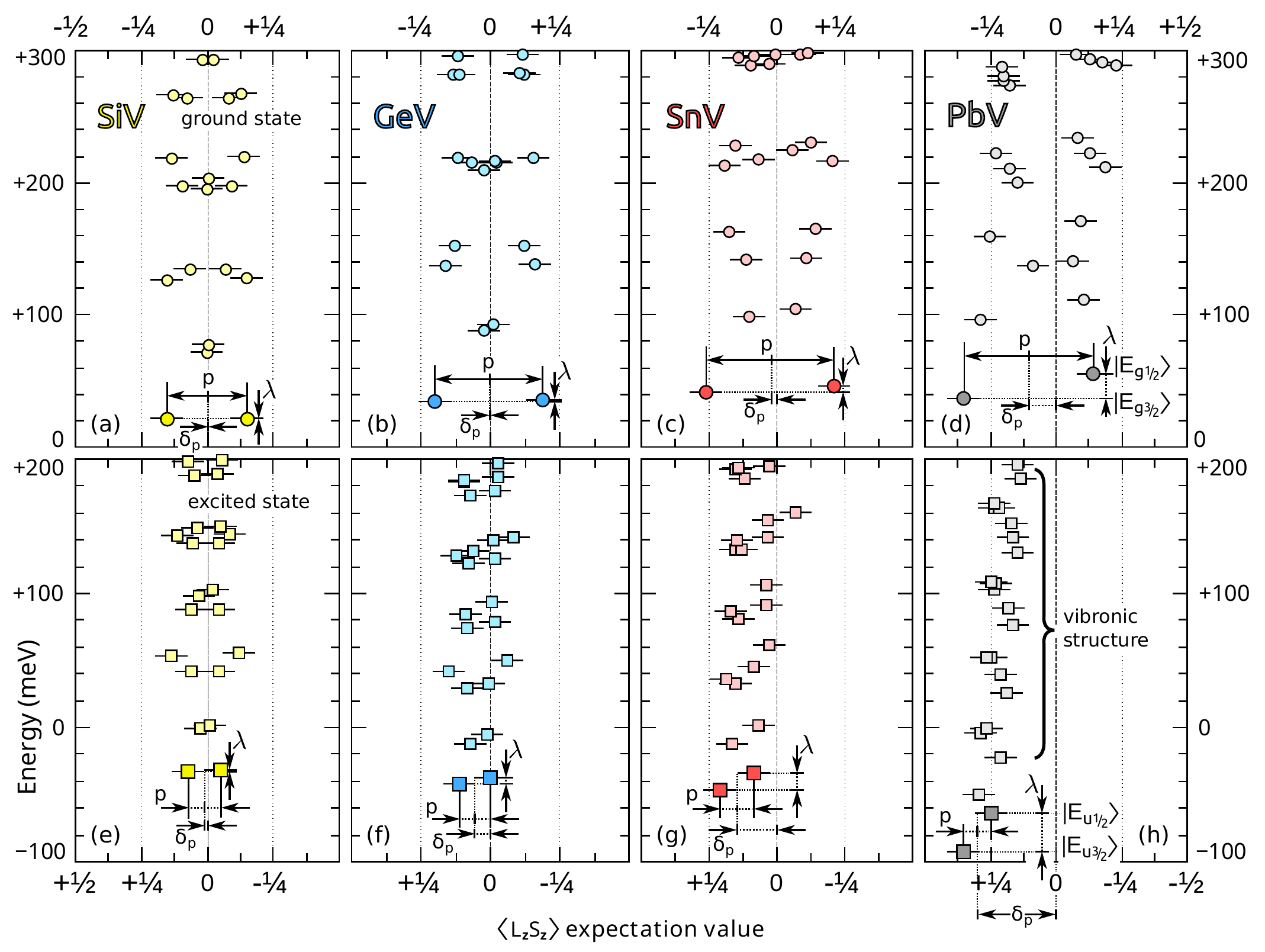} 
	\caption{\label{fig:SOC} 
		Calculated eigenvalues of the adjoint DJT (Eq. \eqref{eq:DJT}) and SOC (Eq. \eqref{eq:SOC}) interaction for the XV($-$) color centers. The two lowest eigenvalues for each figure correspond to the $^2 E_{g,u}$ vibrionic ground states for $g$ ground (a-d) and the optically allowed $u$ excited (e-h) states. All the considered eigenvalues are doubly degenerate in spin dimension because of the Kramers degeneracy. Each state consists of pure 
$\uparrow$ or $\downarrow$ spin state, thus the fourfold degeneracy of $^2E_{g,u}$ is fulfilled. We label the energy difference of the two lowest energy states by $\lambda$ that is directly observed in the fine structure of the ZPL in the PL spectrum known as zero-field-splitting. Along $x$ axis we depict the eigenvalues with respect to their partially quenched spin-orbit coupling strength $\left\langle L_{z}S_{z}\right\rangle $, thus one can directly read out the $p$ factors from this figure. $\delta_p$ shows deviation of the Ham reduction factors on $E_{g,u\nicefrac{3}{2}}$ and $E_{g,u\nicefrac{1}{2}}$ states. The larger the $\delta_p$ the less accurate is the treatment of SOC as a perturbation over DJT. We note that a mirror symmetry at $x=0$ shows up for the ground state of SiV center in the entire vibronic spectrum that demonstrates that SOC can be treated as a perturbation over JT effect. The systematic left-shift at $x$ axis for the vibronic spectrum of SnV and PbV defects implies that SOC is comparable with the JT coupling.}
\end{figure*}

We calculated the $p$ reduction factors for the $^2E_u$ excited state of the 
XV($-$) defects too (see Table~\ref{tab:SOC}). In the excited state, the Ham reduction factors are significantly smaller than those in the groundstate because 
the $E_\text{JT}$ energies are larger in the excited state than those in the groundstate (c.f.\ Table~\ref{tab:SOC}). We think that the larger electron density in the interstitial region around the impurity atom in the $^2E_u$ state contributes to form long bonds between the carbon atoms around the impurity atom and thus makes the Jahn-Teller distorted structure more favorable in that state than in the $^2E_g$ state. Surprisingly the Ham factor $p$ scheme provides reasonable results even for the optically excited state of PbV($-$) for which $\lambda_0 > E_\text{JT}$, nevertheless, exact diagonalization of the adjoint DJT and SOC Hamiltonians is needed for accurate results. We show a graphical interpretation for the spin-orbit and electron-phonon coupled systems in Fig.~\ref{fig:SOC}. We expand the series expansion with the spin degrees of freedom in Eq.~\eqref{eq:DJTseries} then directly diagonalize the Hamiltonian including the SOC and DJT effects simultaneously. As long as $\lambda_0 \ll E_{JT}$ SOC is only a small perturbation over the polaronic DJT groundstate, and the $^2E_{g,u}$ $4\times$ degenerate level is split into double degenerate $E_{g,u\nicefrac{3}{2}}$ and $E_{g,u\nicefrac{1}{2}}$ levels. We label this splitting with $\lambda$ in Fig.~\ref{fig:SOC} that is directly observed in the fine structure of the ZPL optical emission. The perturbative approach of SOC is valid mostly for the SiV($-$) system, especially on its optically ground state (a), thus approximation of $\lambda$ with 
$\lambda_\text{Ham}$ is valid. If SOC energy is higher than $E_\text{JT}$ then Ham reduction scheme still provides surprisingly good results when compared to those from exact diagonalization. Considerable deviations only begin to appear for PbV($-$) (see also Table~\ref{tab:SOC}). For SiV($-$), the plotted data is symmetric respect to the $x=0$ axis as the 
$E_{u\nicefrac{1}{2}}$ is increased by $p \lambda_0 /2$ energy and $E_{u\nicefrac{3}{2}}$ is decreased by the same amount. However, for the excited state of PbV($-$), this is not symmetric. The SOC systematically shifts the eigenvalues to the left in $x$-axis. In this case, the $E_{u\nicefrac{1}{2}}$ state contains larger contribution from spin-orbit favored $e_{u-}^{\downarrow}$ and $e_{u+}^{\uparrow}$ states rather than from the unfavorable $e_{u-}^{\uparrow}$ and $e_{u+}^{\downarrow}$ states, indicating that the DJT and SOC Hamiltonians should be solved simultaneously.

\subsection{Spin Hamiltonian for Group-IV -- Vacancy qubits and its implications}

By applying an external constant magnetic field, the spin double degenerate levels of the XV($-$) color centers may split. Previously, a spin Hamiltonian was deduced for SiV($-$) qubit~\cite{SiV_Hepp2014} to describe this feature that we further develop based on the coupled spin-orbit-phonon Hamiltonian for the groundstate ($g$) and excited state ($u$) as follows
\begin{equation}
\label{eq:Hspin}
\begin{aligned}\!\!\!\!\!\hat{H}_{\text{eff}}^{g,u}= & \underset{{\displaystyle -\lambda^{g,u}}}{\underbrace{-\left(p^{g,u}\lambda_{0}^{g,u}+K_{JT}\right)}}\hat{L}_{z}\hat{S}_{z}+\mu_{B}\underset{{\displaystyle f^{g,u}}}{\underbrace{p^{g,u}g_{L}^{g,u}}}\hat{L}_{z}B_{z}+\\
& +\mu_{B}g_{S}\hat{\boldsymbol{S}}\boldsymbol{B}-\underset{{\displaystyle 2\delta_{f}^{g,u}}}{\underbrace{2\delta_{p}^{g,u}g_{L}^{g,u}}}\hat{S}_{z}B_{z}+\hat{\varUpsilon}_{\mathrm{strain}}\text{,}
\end{aligned}
\end{equation} 
where $g_S=2.0023$ is the $g$-factor of the electron, $\mu_B$ is the Bohr-magneton of the electron, $\boldsymbol{B}$ is the external homogeneous magnetic field, $B_z$ is its $z$ component with $z$-axis parallel to the symmetry axis of the defect. We note that the hypefine interaction between the electron spin and nuclear spins in the diamond lattice or with the impurity atom is not considered here. In the braces we merge the different effects into single effective parameters where we follow the nomenclature of Ref.~\onlinecite{SiV_Hepp2014} for the two common parameters $\lambda$ and $f$, whereas parameter $\delta_f$ appears as a new parameter according to our derivation. Our derivation reveals the microscopic origin of the spin Hamiltonian merged parameters that will be discussed below. The operator $\hat{L}_z$ acts on the $E_{g,u}$ orbitals as $\pm1$, where $g$ and $u$ refers again to the parity of the wavefunctions with $g$ groundstate and $u$ optically allowed excited state. 

The derivation of the terms can be found in Appendix~\ref{sec:dereffHam} and \ref{sec:origStevens}. Here we discuss all the resultant terms in our spin Hamiltonian in details. The first term in Eq.~\eqref{eq:Hspin} contains an effective spin-orbit splitting. We first note the negative sign which originates from the three-electron many-body $E_g$ and $E_u$ states. As a consequence, $E_{g,u\nicefrac{3}{2}}$ is lower in energy than the $E_{g,u\nicefrac{1}{2}}$, in contrast to the previous assignments. Here, we label the states with $m_j$ quantum numbers which is the sum of $m_l=\pm1$ orbital angular momentum and the $m_s=\pm\nicefrac{1}{2}$ spin quantum numbers. Since the orbitals are coupled to phonons the Ham reduction factors $p$ can be different for $\tilde{E}_{g,u\nicefrac{3}{2}}$ and $\tilde{E}_{g,u\nicefrac{1}{2}}$ polaronic wavefunctions (see Eqs.~\eqref{eq:p_ge}), and the final $p^{g,u}$ will be the average of the two (see Eq.~\eqref{eq:Heff_with_braces}). In addition, the vibronic zero-point energy of these states can also differ, in principle, that will change the energy gap between these two states that we label by $K^{g,u}$ as defined in Eq.~\eqref{eq:K}. We note that $K^{g,u}$ can be neglected for SiV($-$) and GeV($-$) but it becomes substantial for SnV($-$) and PbV($-$). 

In the second term two reduction factors appear. The $p^{g,u}$ is already introduced above and caused by electron-phonon coupling. The $g_{L}^{g,u}$ orbital reduction factor was previously discussed by Stevens~\cite{Stevens542}, thus we call it Stevens' orbital reduction factor. This originates from the fact that orbital angular moment $\hat{L}_z$ is only an effective operator as the $D_{3d}$ point group of the XV($-$) systems does not respect the full $O(3)$ rotational symmetry as illustrated in Appendix~\ref{sec:origStevens}. We find that the $g_{L}^{g}$ is in particular substantially smaller than one, and significantly reduces the effective $\hat{L}_z$ (see Appendix~\ref{sec:origStevens}). 

The third term is the usual Zeeman term for electron spins. The fourth term provides a correction to the $g_S$ constant, thus modifies it to a tensor where the corrected $zz$ component is caused by the polaronic nature of the $\tilde{E}_{g,u\nicefrac{3}{2}}$ and $\tilde{E}_{g,u\nicefrac{1}{2}}$ states where the $\delta^{g,u}_p$ scales with the difference of the corresponding Ham reduction factors (see Eqs.~\eqref{eq:delta_ge} and \eqref{eq:Heff_with_braces} in Appendix~\ref{sec:dereffHam}). This is a new correction which is negligible for SiV($-$) but already appears for the excited state of GeV($-$), and both for the ground and excited state of SnV($-$) and PbV($-$).

We propose that the $\hat{\varUpsilon}_{\mathrm{JT}}$ operator in a previous study~\cite{SiV_Hepp2014} that was associated with the Jahn-Teller effect originates from the residual strain around the individual SiV($-$) centers, thus we propose to label as a $\hat{\varUpsilon}_{\mathrm{strain}}$ operator instead. We note that the strain was also considered in that study which can be very strong (in the order of 100~GHz) in nanodiamond samples and much smaller in the bulk diamond samples (few GHz) that is fairly described in Chapters 4.1 and 4.3 in Ref.~\onlinecite{Hepp_dissertation}. We find that the strong electron-phonon coupling occurs for the phonons with the energy in the order of 10~meV that is clearly manifested in the PL spectrum of SiV($-$). This energy region is significantly larger than the spin-orbit energy, thus the strong electron-phonon coupling, i.e., the DJT effect is manifested via Ham reduction of the spin-orbit coupling. We note that the very low energy acoustic $E_g$ phonons might be considered as static strain which would treat these distortions by static Jahn-Teller effect instead of DJT. In any case, the final form of that Hamiltonian is the strain Hamiltonian, thus the two effects cannot be distinguished in experiments. 

\begin{figure*}[ht] \includegraphics[width=\textwidth]{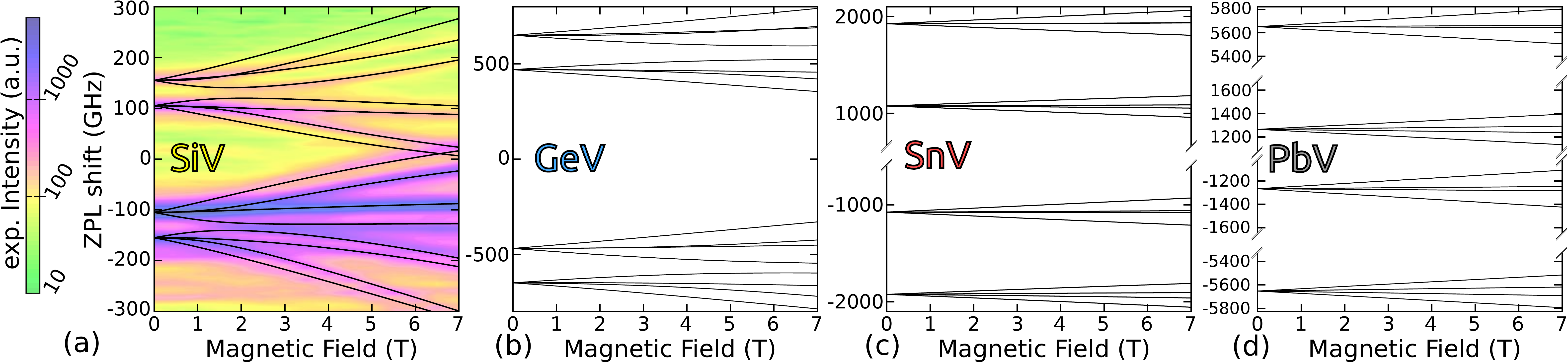} 
\caption{\label{fig:Bdepend} 
Magnetic field dependence of the ZPL splitting where the magnetic field is aligned along $\langle100\rangle$ direction. Zero energy is aligned to no spin-orbit and no-phonon coupling solution. The $\lambda_0$ SOC coupling is fitted to obtain the experimental zero-field-splitting for SiV, GeV, and SnV in order to have a direct comparison to the experimental data whereas full \emph{ab initio} result is shown for PbV. The actual parameters are listed in Table~\ref{tab:SOCDJT} in Appendix \ref{sec:dereffHam} for the effective spin Hamiltonian Eq.~\eqref{eq:Hspin}. We plot all possible transitions between the $^2E_g$ and $^2E_u$ states, thus there are total $4\times4$ individual lines for each figure. The plotted emission intensity in (a) is experimental data extracted from Fig.~4.10 on page~131 in Ref.~\onlinecite{Hepp_dissertation}.}
\end{figure*}

Based on these considerations, we simulate the Zeeman splitting of the corresponding states with a magnetic field aligned to $\langle100\rangle$ direction in Fig.~\ref{fig:Bdepend} where we assume that no strain acts on the XV($-$) color centers. In these simulations, we used \emph{ab initio} electron-phonon deduced parameters but the intrinsic spin-orbit energies ($\lambda_0$) were scaled in order to reproduce the experimental zero-field-splittings ($\lambda$). We followed this procedure in order to directly compare our results to the experimental spectrum for SiV($-$), GeV($-$) and SnV($-$). The $g^{g,u}_L$ parameters were fitted to obtain the experimental spectrum of SiV($-$) and we applied these Stevens' orbital reduction factors for the other XV($-$) qubits. This is a simple approximation that could lead to a slightly underestimated values for the ground state of GeV($-$), SnV($-$) and PbV($-$), where the $d$ orbitals of the impurity atom may contribute to $g^{g}_L$. Nevertheless, our simulations should result in relatively accurate spectra. For SiV($-$), our procedure resulted in $g^g_L=0.328$ and $g^u_L=0.782$. The strong reduction in the ground state can be understood by the shape of the corresponding orbital that we depict in Fig.~\ref{fig:Stevens_reduction} in Appendix~\ref{sec:origStevens}. For instance, $E_{g+}$ state not only transforms as $m_l=+1$ wavefunction but also as $m_l=-2$ wavefunction. The linear combinations of the two leads to a significant reduction in the effective interaction with the external magnetic field, i.e., relatively small 
$g_L$. In the $E_u$ excited state, the impurity $p$ orbitals contribute to the interaction with the magnetic field unlike in the $E_g$ ground state, thus the reduction parameter is significantly larger for the excited state (see Appendix~\ref{sec:origStevens}). The list of all parameters can be found in Table~\ref{tab:SOCDJT} in Appendix~\ref{sec:dereffHam}. For SiV($-$), our spin Hamiltonian can well reproduce the curvatures of the experimental Zeeman spectrum. We apply the same Hamiltonian for the other XV($-$) qubits. Beside the obvious growing ZFS going from smaller to larger atomic number of impurity atoms, the general trend is that the curves are steeper for larger atomic number of impurity atoms because of the enhanced $g_{zz}$ values which is caused by the complex spin-orbit-phonon coupling (the fourth term in Eq.~\eqref{eq:Hspin}).  

For PbV($-$) the calculated zero-field-splitting $\lambda$ in the $^2E_g$ ground state is about 4.4~THz or 18.2~meV which means that the $^2E_{g\nicefrac{3}{2}}$ groundstate will be thermally filled with 100\% and 
$\approx$96\% occupation at cryogenic and liquid nitrogen temperature,
respectively, 
that can be useful for quantum optics protocols. By applying the theory developed for the estimation of the coherence time of the SiV($-$) (see Ref.~\cite{Jahnke2015} and the Supplementary Material in Ref.~\onlinecite{Iwasaki:PRL2017}), we find that the decoherence process caused by the acoustic phonons is completely quenched at cryogenic temperatures for PbV($-$) because of the large $\lambda$ between the two branches of the $^2E_g$ ground state, and the coherence time of the electron spin should be only limited by the nuclear spins or electron spins in the diamond crystal. That is much more
practical for quantum communication applications compared to the
millikelvin cooling needed for similar coherence time in SiV($-$)
qubit~\cite{Sukachev:PRL2017}. We note that the energy gap in SnV($-$) should result in microsecond to millisecond coherence time for the electron spin going from 4~K to 1~K measurement temperature~\cite{Iwasaki:PRL2017}. 
 
\section{Summary and conclusion}
\label{sec:Conclusion}
We performed a systematic study on the magneto-optical properties of 
Group-IV -- Vacancy color centers in diamond by means of \emph{ab initio} density functional theory calculations. We identified the photostability of these centers that can act as solid state qubits. We developed a novel spin Hamiltonian for these qubits in which the electron angular momentum and spin as well as the phonons are strongly coupled and identified such terms that have not been considered so far but are important in understanding their magneto-optical properties. We solved \emph{ab initio} this complex problem for the model of these color centers consisting of up to 1000-atom supercells, and were able to reproduce previous experimental data. Furthermore, we identified SnV($-$) and PbV($-$) qubits with long spin coherence time at cryogenic temperatures where the spin state of PbV($-$) can also be thermally initialized at these temperatures. Our \emph{ab initio} toolkit and spin Hamiltonian analysis serve as a template for similar studies in 3D materials such as silicon carbide or 2D materials such as hexagonal boron nitride or transitional metal dichalgonides (TMD) or dioxides (TMO) which are fast emerging materials hosting qubits or single photon sources. In particular, the TMD and TMO materials exhibit strong spin-orbit couplings induced by the transition metal ions in the crystal in which strong mixing of spin-orbit and electron-phonon coupling are expected in the defects acting as qubits, and they should be treated at equal footing.

\section*{Acknowledgments} 

Support \'UNKP-17-3-III New National Excellence Program of the Ministry of Human Capacities of Hungary, the National Research Development and Innovation Office of Hungary within the Quantum Technology National Excellence Program 
(project contract No.~2017-1.2.1-NKP-2017-00001)
and the EU Commission (DIADEMS Project Contract No.~611143)
is acknowledged.



\appendix
\section{Convergence of spin-orbit energies as a function of supercell size}
\label{sec:SOCscale}
\begin{figure*}
\includegraphics[width=1\textwidth]{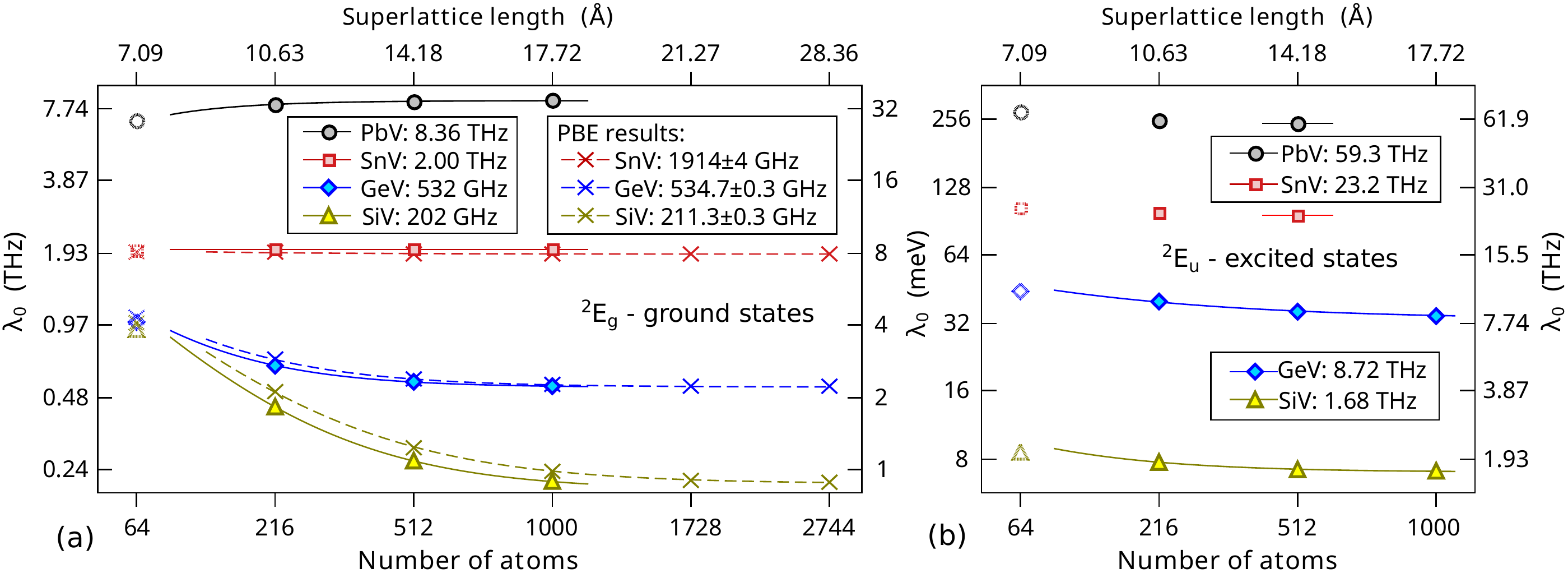}
\caption{\label{fig:SOCscale}Calculated spin-orbit energies with HSE06 DFT functional as a function of the size of the supercell within $\Gamma$-point approximation. The converged values are given in the label boxes. We show the computationally less expensive PBE results for giant supercells too. The 5 data points for PBE yields a standard error for the fitting. The small standard error in these calculations supports the exponential decay for HS06 data points.}
\end{figure*}

Here we provide the calculated spin-orbit energies for the ground ($g$) and excited ($u$) state as a function of the size of the supercell within $\Gamma$-point sampling of the Brillouin-zone by HSE06 DFT functional. We determined the $\lambda_0$ intrinsic SOC parameters by fitting the $\lambda_0(L) = \lambda_0 + A \exp \left( - L \cdot B  \right)$ function on the data points as obtained from 216, 512 and 1000-atom diamond supercells (see Fig.~\ref{fig:SOCscale}).  Here $L$ is the length of the corresponding cubic diamond supercell, and $\lambda_0$, $A$, 
and $B$ are the fitting parameters. We also calculated the effective SOC for the optically active $^2E_u$ excited states where the calculation procedure goes as described for the $E_g$ state in Sec.~\ref{sec:Methods}. We note that the 512-atom supercell is proven to be convergent for $^2E_u$ spin-orbit energies in 
SnV($-$) and PbV($-$) defects, thus we use those energies as isolated converged values. We prove the exponential decay of SOC parameters by the computationally less demanding Perdew-Burke-Ernzerhof\cite{PBE} (PBE) calculations  that allowed us to use giant supercells in the scaling study (see Fig.~\ref{fig:SOCscale}).

\section{Derivation of the Ham reduction factor}
\label{sec:derHamfactor}

Here we briefly derive the Ham reduction factor for the cases where the spin-orbit coupling is significantly smaller than the electron-phonon interaction, and it can be treated as a first order perturbation. We start with the definition of the electronic orbitals either in the real or imaginary forms,
\begin{equation}
\left|e_{g\pm}\right\rangle =\frac{1}{\sqrt{2}}\left(\left|e_{x}\right\rangle \pm i\left|e_{y}\right\rangle \right)=\frac{1}{\sqrt{2}}\left(\begin{array}{c}
1\\
\pm i
\end{array}\right)\begin{array}{c}
\leftarrow|e_{x}\rangle\\
\leftarrow|e_{y}\rangle
\end{array} \text{.}
\end{equation}
The vibronic wavefunction $\tilde{\Psi}_{\pm}$ caused by electron-phonon interaction can be expanded as written in Eq.~\eqref{eq:DJTseries} that mixes the electronic orbitals and $|n,m\rangle$ $e_g$ phonons. The spin-orbit coupling then should be
calculated for the $\tilde{\Psi}_{\pm}$ wavefunction, 
\begin{equation}
\label{eq:Hsoc}
\ensuremath{\lambda_{\mathrm{Ham}}=2\left|\left\langle \tilde{\Psi}_{\pm}\right|\hat{H}_\text{SOC}\left|\tilde{\Psi}_{\pm}\right\rangle \right|}\text{.}
\end{equation}
By expanding $\tilde{\Psi}_{\pm}$ in Eq.~\eqref{eq:Hsoc} we arrive at 
\begin{widetext}
\begin{equation}
\ensuremath{=\sum_{n,m,k,l}\left[c_{nm}\left\langle e_{g\pm}\right|\left\langle n,m\right|+d_{nm}\left\langle e_{g\mp}\right|\left\langle n,m\right|\right]\hat{H}_\text{SOC}\left[c_{kl}\left|e_{g\pm}\right\rangle \left|k,l\right\rangle +d_{kl}\left|e_{g\mp}\right\rangle \left|k,l\right\rangle \right]}
\end{equation}
\end{widetext}
which leads to
\begin{widetext}
\begin{equation}
\ensuremath{\lambda_{0}\sum_{n,m}\Biggl[c_{nm}^{2}\underbrace{\left\langle e_{g\pm}\right|\hat{\sigma}_{y}\left|e_{g\pm}\right\rangle }_{\underbrace{\frac{1}{2}\left(\!\!\begin{array}{cc}
1 & \!\!\mp i\!\!\end{array}\right)\left(\begin{array}{cc}
\!\! & \!\!\!\negthinspace-i\!\!\\
\!\!i & \!\!\!\!\!\!
\end{array}\right)\left(\begin{array}{c}
1\!\!\\
\!\!\pm i\!\!
\end{array}\right)}_{{\textstyle =+1}}}+ d_{nm}^{2}\underbrace{\left\langle e_{g\mp}\right|\hat{\sigma}_{y}\left|e_{g\mp}\right\rangle }_{\underbrace{\frac{1}{2}\left(\begin{array}{cc}
\!\!1 & \!\!\pm i\end{array}\!\!\right)\left(\begin{array}{cc}
\!\! & \!\!\!\!-i\!\!\\
\!\!i & \!\!\!\!\!\!
\end{array}\right)\left(\begin{array}{c}
\!\!1\!\!\\
\!\!\!\!\mp i\!\!
\end{array}\right)}_{{\textstyle =-1}}}+2c_{nm}d_{nm}\underbrace{\left\langle e_{g\pm}\right|\hat{\sigma}_{y}\left|e_{g\mp}\right\rangle }_{{\textstyle =0}}\Biggr]}
\end{equation}
\end{widetext}
that is 
\begin{equation}
\lambda_{0}\sum_{n,m} \Biggl[c_{nm}^{2}-d_{nm}^{2}\Biggr] = \lambda_{0}p\text{.}
\end{equation}
That final equation defines Ham reduction factor $p$.

\section{Derivation of the effective Hamiltonian}
\label{sec:dereffHam}
In Eq.~\eqref{eq:DJTseries} the lower-energy ($\nicefrac{3}{2}$) and higher-energy ($\nicefrac{1}{2}$) states split by spin-orbit interaction shared the same 
$p$ reduction factors in their corresponding ground and excited state. However, the $\left\langle L_{z}S_{z}\right\rangle$ weights are not symmetric as shown in 
Fig.~\ref{fig:SOC} due to the electron-phonon coupling, thus we introduce individual Hamiltonian for each doublet ($\Gamma=E_{g\nicefrac{1}{2}}$, $E_{g\nicefrac{3}{2}}$, $E_{u\nicefrac{1}{2}}$, $E_{u\nicefrac{3}{2}}$) in the following, 
\begin{equation}
\hat{H}^{\Gamma}_{\text{eff}}=\left|\Gamma\right\rangle \left\langle \Gamma\right|\left[p_{\Gamma}^{g,u}\hat{L}_{z}(-\lambda_{0}^{g,u}\hat{S}_{z}+g_{L}^{g,u}B_{z})+g_{S}\hat{\boldsymbol{S}}\boldsymbol{B}\right]\text{,}
\label{eq:Hgamma}
\end{equation}
where $\left|\Gamma\right\rangle \left\langle \Gamma\right|$
projector ensures that we stay in the $\Gamma$ doublet in Eq.~\ref{eq:Hgamma}. The projectors can be expressed by spin-orbit operators as
\begin{subequations}
\begin{gather}
\left|E_{g,u\nicefrac{3}{2}}\right\rangle \left\langle E_{g,u\nicefrac{3}{2}}\right|=\frac{1}{2}+\hat{L}_{z}\hat{S}_{z}
\\
\left|E_{g,u\nicefrac{1}{2}}\right\rangle \left\langle E_{g,u\nicefrac{1}{2}}\right|=\frac{1}{2}-\hat{L}_{z}\hat{S}_{z} \text{.}
\end{gather}
\label{eq:spin-orbit}
\end{subequations} 
We note that as $\left|\Gamma\right\rangle \left\langle \Gamma\right|$ commutes with other parts of $\hat{H}^{\Gamma}$, one needs to include the projector only once in Eq.~\eqref{eq:Hgamma}. The actual $p_{\Gamma}$ values can be directly read out from Fig.~\ref{fig:SOC} that we define as 
\begin{subequations}
\begin{gather}
p_{\nicefrac{3}{2}}^{g,u}=+2\left\langle \tilde{E}_{g,u\nicefrac{3}{2}}\right|\hat{L}_{z}\hat{S}_{z}\left|\tilde{E}_{g,u\nicefrac{3}{2}}\right\rangle 
\\
p_{\nicefrac{1}{2}}^{g,u}=-2\left\langle \tilde{E}_{g,u\nicefrac{1}{2}}\right|\hat{L}_{z}\hat{S}_{z}\left|\tilde{E}_{g,u\nicefrac{1}{2}}\right\rangle \text{.} 
\end{gather}
\label{eq:p32_12}
\end{subequations}
These parameters are the expectation value of the (dimensionless)
SOC acted on the electron-phonon coupled wavefunctions. The $\tilde{E}_{g,u\nicefrac{1}{2}}$ and $\tilde{E}_{g,u\nicefrac{3}{2}}$ are the wavefunctions from series expansion of Eq.~\eqref{eq:DJTSOCseries} of the adjoint SOC and DJT interaction. In the next step, we unify the $\nicefrac{3}{2}$ and $\nicefrac{1}{2}$ substates inside the ground and excited state manifold,
\begin{widetext}
\begin{equation}
\hat{H}_{\text{eff}}^{g,u}=\!\!\!\!\!\!\!\!\sum_{\Gamma=E_{g,u\nicefrac{1}{2}},E_{g,u\nicefrac{3}{2}}}\!\!\!\!\!\!\hat{H}_{\text{eff}}^{\Gamma}=\left(\frac{1}{2}-\hat{L}_{z}\hat{S}_{z}\right)p_{\nicefrac{1}{2}}^{g,u}\hat{L}_{z}(-\lambda_{0g,u}\hat{S}_{z}+g_{L}^{g,u}B_{z})+\left(\frac{1}{2}+\hat{L}_{z}\hat{S}_{z}\right)p_{\nicefrac{3}{2}}^{g,u}\hat{L}_{z}(-\lambda_{0}^{g,u}\hat{S}_{z}+g_{L}^{g,u}B_{z})+g_{S}\hat{\boldsymbol{S}}\boldsymbol{B} \text{,}
\label{eq:Heff_with_p}
\end{equation}
where we can take the advantage of the properties of $\hat{L}_z$ and $\hat{S}_z$ to arrive at
\begin{equation}
\hat{H}_{\text{eff}}^{g,u}=\underset{=p_{g,u}}{\underbrace{\frac{1}{2}\left(p_{\nicefrac{3}{2}}^{g,u}+p_{\nicefrac{1}{2}}^{g,u}\right)}}\hat{L}_{z}\left(-\lambda_{0g,u}\hat{S}_{z}+g_{L}^{g,u}B_{z}\right)+\underset{\text{offset}}{\underbrace{\left(p_{\nicefrac{3}{2}}^{g,u}-p_{\nicefrac{1}{2}}^{g,u}\right)\lambda_{0}^{g,u}}}\underset{\nicefrac{1}{4}}{\underbrace{\hat{L}_{z}^{2}\hat{S}_{z}^{2}}}-\underset{=2\delta_{g,u}^{f}}{\underbrace{\left(p_{\nicefrac{3}{2}}^{g,u}-p_{\nicefrac{1}{2}}^{g,u}\right)g_{L}^{g,u}}}\hat{S}_{z}B_{z}\underset{1}{\underbrace{L_{z}^{2}}}+g_{S}\hat{\boldsymbol{S}}\boldsymbol{B} \text{.}
\label{eq:Heff_with_braces}
\end{equation}
\end{widetext}
Here we substitute $\hat{L}_{z}^{2}=1$ and $\hat{S}_{z}^{2}=\frac{1}{4}$ for XV($-$) systems. The offset term in Eq.~\eqref{eq:Heff_with_braces} does not depend on the magnetic field or cause splitting between the branches of the ground and excited state, therefore it is not observable in the experiments. That is rather a correction term for the first principles calculations after turning on the spin-orbit and electron-phonon couplings. We use this correction in the calculation of the ZPL energies. Next, we only consider the observable terms in the spin Hamiltonian derivation for XV($-$) color centers which goes as 
\begin{equation}
\hat{H}_{\text{eff}}^{g,u}=-\lambda^{g,u}L_{z}S_{z}+f^{g,u}L_{z}B_{z}+g_{S}\boldsymbol{SB}-2\delta_{f}^{g,u}S_{z}B_{z}\text{,}
\label{eq:Heff}
\end{equation} 
where we introduced the following parameters,
\begin{equation}
\lambda^{g,u}=p^{g,u}\lambda^{g,u}_0+K_{JT}^{g,u}\quad f^{g,u}=p^{g,u}g^{g,u}_{L}\quad\delta^{g,u}_{f}=\delta^{g,u}_{p}g^{g,u}_{L}\text{.}
\label{eq:lambda_ge}
\end{equation} 

The actual parameters can be calculated from the DJT and SOC entangled wavefunctions, i.e., the $E_{g,u\nicefrac{3}{2}}$ and 
$ E_{g,u\nicefrac{1}{2}}$ doublets as
\begin{equation}
\begin{aligned}p^{g,u}=\frac{p_{\nicefrac{3}{2}}^{g,u}+p_{\nicefrac{1}{2}}^{g,u}}{2}=\left\langle \tilde{E}_{g,u\nicefrac{3}{2}}\right|\hat{L}_{z}\hat{S}_{z}\left|\tilde{E}_{g,u\nicefrac{3}{2}}\right\rangle +\\
\left\langle \tilde{E}_{g,u\nicefrac{1}{2}}\right|\hat{L}_{z}\hat{S}_{z}\left|\tilde{E}_{g,u\nicefrac{1}{2}}\right\rangle 
\end{aligned}
\label{eq:p_ge}
\end{equation}
and
\begin{equation}
\begin{aligned}\delta_{p}^{g,u}=\frac{p_{\nicefrac{3}{2}}^{g,u}-p_{\nicefrac{1}{2}}^{g,e}}{2}=\left\langle \tilde{E}_{g,u\nicefrac{3}{2}}\right|\hat{L}_{z}\hat{S}_{z}\left|\tilde{E}_{g,u\nicefrac{3}{2}}\right\rangle -\\
\left\langle \tilde{E}_{g,u\nicefrac{1}{2}}\right|\hat{L}_{z}\hat{S}_{z}\left|\tilde{E}_{g,u\nicefrac{1}{2}}\right\rangle \text{.}
\end{aligned}
\label{eq:delta_ge}
\end{equation}

The term $K_{JT}^{g,u}$ requires further explanation. This comes from the fact that the zero-point energy of the vibronic $^{2}E_{g,u\nicefrac{3}{2}}$ and $^{2}E_{g,u\nicefrac{1}{2}}$ wavefunctions are not the same, thus this can cause an extra splitting labeled by $K_{JT}^{g,u}$ that adds to $p^{g,u}\lambda^{g,u}$.
$K_{JT}^{g,u}$ can be calculated as
\begin{equation}
\begin{aligned}K_{JT}^{g,u}=\left\langle \tilde{E}_{g,u\nicefrac{1}{2}}\right|\hat{H}_{DJT}\left|\tilde{E}_{g,u\nicefrac{1}{2}}\right\rangle +\\
-\left\langle \tilde{E}_{g,u\nicefrac{3}{2}}\right|\hat{H}_{DJT}\left|\tilde{E}_{g,u\nicefrac{3}{2}}\right\rangle \text{.} 
\end{aligned}
\label{eq:K}
\end{equation}
This term gives negligible correction to the ZFS of GeV($-$) and SiV($-$) but becomes significant for PbV($-$) as shown in Table~\ref{tab:SOCDJT}.
\begin{table*}[] 
\caption{\label{tab:SOCDJT} Proposed parameters of the effective Hamiltonian in Eq.~\eqref{eq:Hspin} for XV($-$) systems. $\lambda_0$ is scaled to reproduce the experimental $\lambda$ for SiV($-$), GeV($-$) and SnV($-$). The $p$, 
$\delta_p$ reduction parameters are calculated \emph{ab initio} from the SOC and DJT coupled Hamiltonian (see Fig.~\ref{fig:SOC} for visual interpretation). The orbital $g_L$ reduction factors were fit to reproduce the experimental curves for SiV($-$), and the same $g_L$ factors were applied to the other XV($-$) systems. The definition of the parameters shown here can be found in Eqs.~\eqref{eq:p_ge}-\eqref{eq:K}.}
\begin{ruledtabular}
\begin{tabular}{lcccccc ccccc}
			system	        & $\!\!\!\!\!\lambda_0$ (THz)&$p_{\nicefrac{3}{2}}$ & $p_{\nicefrac{1}{2}}$ & $p$ &$\delta_p$&$g_L$&$K_{JT}$ (GHz)&$p\lambda_0$ (GHz) & $\lambda$ (GHz)                                                 & $f$ & $\delta_f$ \\ \hline 
			SiV$(^2E_g)$   	&0.163                      &0.311&0.305&0.308&0.003     &0.328&0.06 &50  &50\footnote{Ref.~\onlinecite{SiV_Hepp2014}   \label{foot:Si2}}        &0.1\textsuperscript{\ref{foot:Si2}}& 0.001 \\	     	
			GeV$(^2E_g)$   	&0.465                      &0.400&0.380&0.390&0.010     &0.328&0.11 &181 &181\footnote{Ref.~\onlinecite{Ekimov:2015GeV} \label{foot:Ge2}}       &0.128&0.003  \\ 	     	
			SnV$(^2E_g)$   	&1.801                      &0.513&0.429&0.471&0.042     &0.328&1.82 &848 &850\footnote{Ref.~\onlinecite{Iwasaki:PRL2017} \label{foot:Sn2}}       &0.154&0.014  \\ 		    
			PbV$(^2E_g)$    &8.361                      &0.705&0.284&0.494&0.211     &0.328&252  &4133&4385\footnote{present \emph{ab initio} result \label{foot:Pb2}}  &0.162&0.069  \\ \hline 		 
			SiV$(^2E_u)$  	&2.034                      &0.156&0.100&0.128&0.028     &0.782&0.39 &260 &260\textsuperscript{\ref{foot:Si2}}                                   &0.1\textsuperscript{\ref{foot:Si2}}&0.022\\	     	
			GeV$(^2E_u)$   	&9.877                      &0.241&-0.014&0.113&0.128     &0.782&-2.91&1123&1120\textsuperscript{\ref{foot:Ge2}}                                  &0.089&0.100  \\ 	     	
			SnV$(^2E_u)$    &23.77                      &0.429&-0.178&0.125&0.303     &0.782&18.0 &2981&3000\textsuperscript{\ref{foot:Sn2}}                                  &0.098&0.238  \\ 		    
			PbV$(^2E_u)$    &59.30                      &0.709&-0.500&0.105&0.604     &0.782&689  &6231&6920\textsuperscript{\ref{foot:Pb2}}                                  &0.082&0.473  \\ 		     	     	 	
		\end{tabular} 
	\end{ruledtabular} 
\end{table*}

\section{Origin of Stevens' orbital reduction factor}
\label{sec:origStevens}

\begin{figure}[ht] 
\includegraphics[width=\columnwidth]{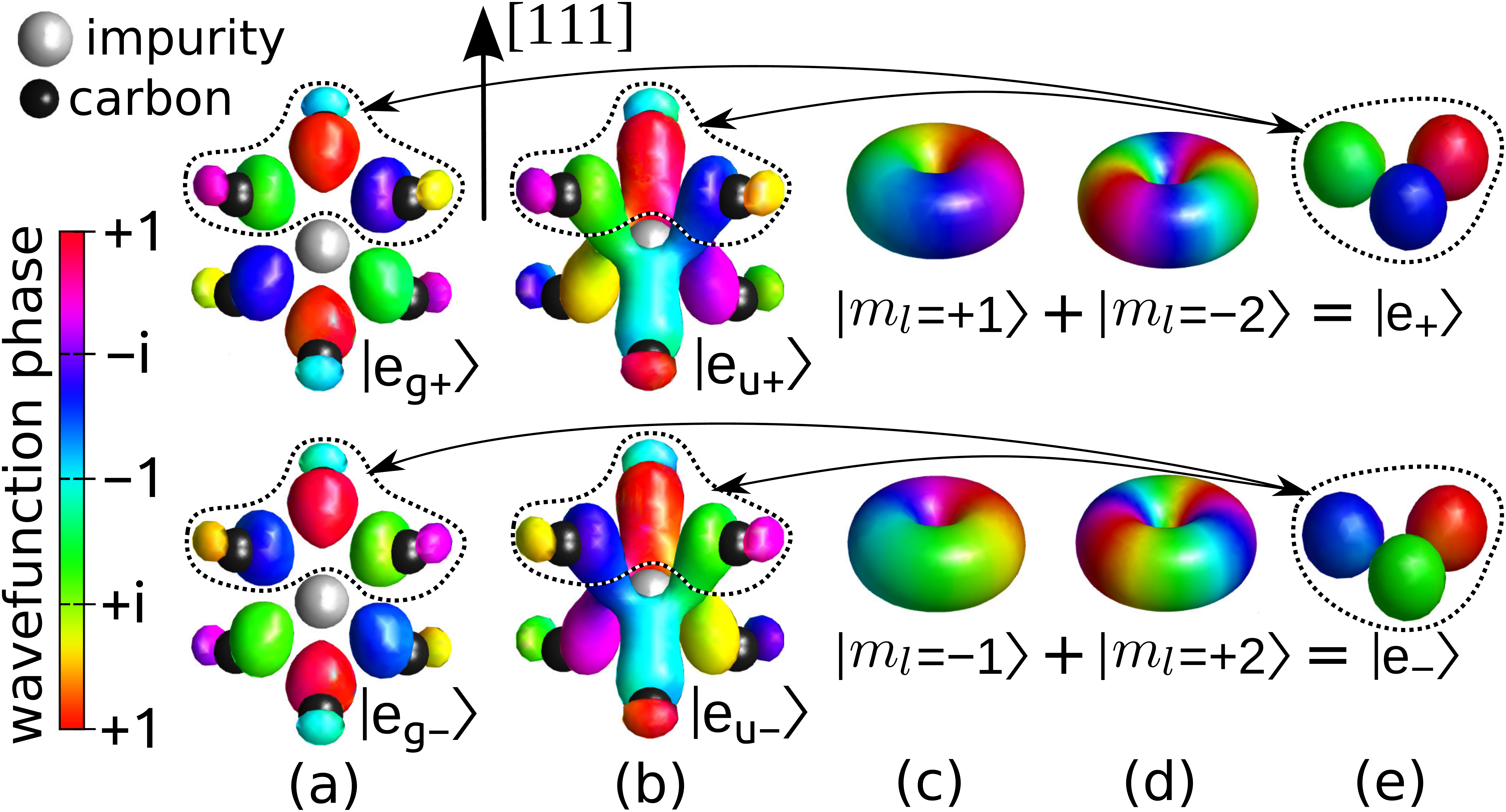} 
\caption{\label{fig:Stevens_reduction} 
Isosurfaces represent the phase of (a) $e_{g\pm}$ orbitals, (b) $e_{u\pm}$ orbitals of SiV($-$) qubit from side view ($z$-axis parallel to [111] direction), (c) $m_l=\pm1$ wavefunctions, and (d) $m_l=\mp2$ wavefunctions. Linear combinations of (c) + (d) (admixtures of atomic $n=2$ $l=1$ $m_l=\pm1$ and $n=3$ $l=2$ $m_l=\mp2$ orbitals) can result in (e) $e_+$ or $e_-$ orbitals very similar to that obtained in (a) and (b). The sign of the wavefunction does not change for ($g$) orbitals whereas it changes for ($u$) orbitals upon inversion.}
\end{figure}

The orbital angular momentum of an atomic orbital may be reduced in the potential created by surrounding ions that reduce the spherical symmetry~\cite{Stevens542}. We show that similar quenching can take place for vacancy-type defects in diamond. In the particular case of XV($-$) color centers, the carbon dangling bonds in the vacancies form double degenerate $E_{g\pm}$ and $E_{u\pm}$ states.  The $E_{g,u\pm}$ states were considered as $m_l=\pm1$ states~\cite{SiV_Hepp2014, Hepp_dissertation} which led to the assumption of $g^{g,u}_{L}=\pm1$. However, one should notice that the $E_{g,u}$ state will also transform as $m_l=\mp2$ under $D_{3d}$ symmetry. We illustrate this by plotting the $e_{g,u\pm}$ orbitals in comparison to the $m_l=\pm1$ and $m_l=\mp2$ wavefunctions in Fig.~\ref{fig:Stevens_reduction}. It can be observed that the $e_{g,u\pm}$ orbitals can be rather described as $\alpha |m_l=\pm1\rangle + \sqrt{1-\alpha^2} |m_l=\mp2\rangle$ where $\alpha$ is a coefficient of the $m_l=\pm1$ contribution. This results in a $|g_{L}|$ that is smaller than 1. Since the $p$ orbitals of the Group-IV atom can contribute to $m_l=\pm1$ in the $E_{u}$ excited state, in contrast to the case of $E_{g}$ ground state, $g^g_L$ is smaller than $g^u_L$ in XV($-$) qubits. We note that the full \emph{ab initio} calculation of the expectation value of $\hat{L}_z$ is not straightforward. The implementation of spin-orbit coupling in \textsc{vasp} is based on the assumption that the 
spin-orbit interaction is predominant close to the core of the atoms and it is negligible in the interstitial regions, which is considered to be a well-justified approximation. In that case, the spin-orbit integrals can be distributed to the spherical volumes (PAW spheres) around the ions with using the atomic wavefunction projectors and potential. However, as Fig.~\ref{fig:Stevens_reduction} demonstrates, the interstitial regions can significantly contribute to the expectation value of 
$\hat{L}_z$, which gives the interaction between the electron wavefunction and the external magnetic field. Therefore, we rather did not estimate this quantity from the values in the PAW sphere of the ions.   


%

\end{document}